\documentclass{article}

\usepackage{microtype}
\usepackage{graphicx}
\usepackage{subfigure}
\usepackage{array}
\usepackage{booktabs} % for professional tables

\usepackage{hyperref}

\usepackage[accepted]{icml2026}

\usepackage{amsmath}
\usepackage{amssymb}
\usepackage{mathtools}
\usepackage{amsthm}

\usepackage[capitalize,noabbrev]{cleveref}

\theoremstyle{plain}

\theoremstyle{definition}

\theoremstyle{remark}

\usepackage[textsize=tiny]{todonotes}

\icmltitlerunning{To What Extent Do Token-Level Representations from Pathology Foundation
Models Improve Dense Prediction?}

\begin{document}

\twocolumn[
\icmltitle{To What Extent Do Token-Level Representations from Pathology Foundation Models Improve Dense Prediction?}

% unless you've provided the [accepted] option to the icml2026


\icmlsetsymbol{equal}{*}
\icmlsetsymbol{cor}{\#}

\begin{icmlauthorlist}
\icmlauthor{Weiming Chen}{equal,yyy}
\icmlauthor{Xitong Ling}{equal,yyy}
\icmlauthor{Xidong Wang}{comp}
\icmlauthor{Zhenyang Cai}{comp}
\icmlauthor{Yijia Guo}{sch}
\icmlauthor{Mingxi Fu}{yyy}
\icmlauthor{Ziyi Zeng}{comp}
\icmlauthor{Minxi Ouyang}{yyy}
\icmlauthor{Jiawen Li}{yyy}
\icmlauthor{Yizhi Wang}{yyy}
\icmlauthor{Tian Guan}{yyy}
\icmlauthor{Benyou Wang}{cor,comp}
\icmlauthor{Yonghong He}{cor,yyy}
\end{icmlauthorlist}

\icmlaffiliation{yyy}{Tsinghua University, Shenzhen, China}
\icmlaffiliation{comp}{CUHK, Shenzhen, China}
\icmlaffiliation{sch}{Peking University, Beijing, China}

\icmlcorrespondingauthor{Yonghong He}{heyh@sz.tsinghua.edu.cn}
\icmlcorrespondingauthor{Benyou Wang}{wangbenyou@cuhk.edu.cn}

\icmlkeywords{Machine Learning, ICML}

\vskip 0.3in
]



\printAffiliationsAndNotice{\icmlEqualContribution} % otherwise use the standard text.

\begin{abstract}
Pathology foundation models (PFMs) have rapidly advanced and are becoming a common backbone for downstream clinical tasks, offering strong transferability across tissues and institutions. However, for dense prediction (e.g., segmentation), practical deployment still lacks a clear, reproducible understanding of how different PFMs behave across datasets and how adaptation choices affect performance and stability. We present PFM-DenseBench, a large-scale benchmark for dense pathology prediction, evaluating 17 PFMs across 18 public segmentation datasets. Under a unified protocol, we systematically assess PFMs with multiple adaptation and fine-tuning strategies, and derive insightful, practice-oriented findings on when and why different PFMs and tuning choices succeed or fail across heterogeneous datasets. We release containers, configs, and dataset cards to enable reproducible evaluation and informed PFM selection for real-world dense pathology tasks. Project Website: \href{https://m4a1tastegood.github.io/PFM-DenseBench}{\textcolor{magenta}{https://m4a1tastegood.github.io/PFM-DenseBench}}.

\end{abstract}

\section{Introduction}
The field of Computational Pathology (CPath) is witnessing a paradigm shift driven by the emergence of Pathology Foundation Models (PFMs). By leveraging self-supervised learning (SSL) on massive archives of whole-slide images (WSIs), models such as UNI \cite{chen2024towards}, Virchow \cite{vorontsov2024foundation}, and Gigapath \cite{xu2024whole} have demonstrated remarkable few-shot and zero-shot capabilities. However, the prevailing evaluation protocols for these models have been predominantly confined to image-level tasks, such as cancer subtyping and slide-level triaging. 

This focus leaves a critical gap in our understanding: while global semantic aggregation is sufficient for classification, clinical diagnosis fundamentally relies on \textbf{dense prediction}—the precise delineation of nuclei, glands, and tissue compartments to quantify tumor microenvironments (TME) and grade malignancies. Despite the central role of segmentation in clinical workflows, it remains unclear whether the "emergent intelligence" observed in classification transfers effectively to pixel-level tasks. Furthermore, the community lacks a systematic understanding of whether standard scaling laws apply to the high-frequency feature extraction required for segmentation, or if the current architectural hegemony of Vision Transformers (ViTs) is truly optimal for delineating fine-grained boundaries. 

% To address these challenges, we present \textbf{PFM-DenseBench}, a comprehensive evaluation framework integrating 18 diverse datasets and 17 state-of-the-art PFMs. We aim to rigorously scrutinize the transferability of foundation models to dense prediction tasks by answering four fundamental research questions: 
To address these challenges, we present \textbf{PFM-DenseBench}, a evaluation framework integrating 18 datasets and 17 PFMs. We aim to scrutinize the transferability of foundation models to dense prediction tasks by answering four research questions:

\begin{itemize}
    \item \textbf{RQ1: Feasibility and Performance Gains.} Can PFMs achieve the same magnitude of performance improvement in dense prediction as observed in classification tasks? Specifically, do pre-trained representations offer a significant advantage over robust, task-specific supervised baselines (e.g., UNet \cite{ronneberger2015u})?
    \item \textbf{RQ2: Granularity-Dependent Generalization.} Does the transferability of PFMs vary across biological scales? We hypothesize that representations optimized for slide-level semantics may behave differently when transferred to nuclei-level (microscopic), gland-level (mesoscopic), and tissue-level (macroscopic) segmentation.
    \item \textbf{RQ3: The Scaling Laws of Dense Prediction.} How do factors such as pre-training data volume, model parameter size, and adaptation capacity (e.g., LoRA rank \cite{hu2022lora}) influence downstream performance? We seek to elucidate whether scaling laws hold true for dense pathological tasks, or if performance saturates differently when precise localization is required.
    \item \textbf{RQ4: Architectural Inductive Biases.} Is the pure Vision Transformer (ViT) paradigm optimal for dense prediction? Given that segmentation relies on local texture and boundary cues, we investigate whether injecting convolutional inductive biases via hybrid adapters offers superior performance.
\end{itemize}
By systematically investigating these questions, this work not only benchmarks the current state of PFMs but also charts a roadmap for the next generation of architecture design in computational pathology.

\section{Related Work}
\paragraph{Pathology Foundation Models.}
Pathology foundation models have substantially advanced computational pathology by providing transferable representations that generalize across diverse downstream tasks. Existing PFMs largely follow two trajectories. The first is vision-only self-supervised pretraining, typically using ViT backbones and objectives such as DINOv2 \cite{oquab2023dinov2}, iBOT \cite{zhou2021ibot}, or masked image modeling \cite{he2022masked} on large pathology corpora (predominantly H\&E, sometimes extended to IHC). The pretrained encoders are then used as generic feature extractors and paired with MIL-style \cite{ling2024agent} \cite{lu2021data} aggregators for WSI-level prediction across classification, retrieval and survival analysis. Representative models include UNI/UNI2 \cite{chen2024towards}, Virchow/Virchow2 \cite{vorontsov2024foundation} \cite{zimmermann2024virchow2}, PathOrchestra \cite{yan2025pathorchestra}, Phikon (v1/v2) \cite{Filiot2023ScalingSSLforHistoWithMIM} \cite{filiot2024phikon}, Prov-GigaPath \cite{xu2024whole}, H-Optimus (v0/v1), Hibou \cite{nechaev2024hibou}, Lunit \cite{kang2022benchmarking}, Kaiko \cite{aben2024towards}, Digepath \cite{zhu2025subspecialty}, StainNet \cite{li2025stainnet} and Midnight-12k \cite{KDK2025}. The second trajectory is \emph{vision--language} pretraining, which leverages textual supervision (e.g., biomedical literature or pathology reports) to learn aligned image--text representations, enabling language-conditioned retrieval, interactive assistance, and report generation; CONCH (v1/v1.5) \cite{lu2024visual} \cite{ding2025multimodal} and MUSK \cite{xiang2025vision} are notable examples. Overall, PFMs are evolving in parallel along robust vision-only representations and semantically grounded multimodal alignment, with increasing emphasis on efficient adaptation to heterogeneous pathology tasks via lightweight, plug-and-play modules.

\paragraph{Dense Prediction.}
Dense prediction is essential in computational pathology because it supports pixel-/region-level quantification of tissue components, enabling clinically relevant measurements (e.g., tumor/stroma/necrosis/immune fractions) and systematic characterization of the tumor microenvironment (TME) \cite{liu2024panoptic}. Most work frames dense prediction as semantic segmentation and adapts strong architectures from natural-image segmentation. TransUNet \cite{chen2021transunet} combines CNN-based local feature extraction with Transformer-based global context modeling to capture both fine morphology and long-range dependencies. ViT-Adapter \cite{chen2022vision} introduces lightweight adaptation modules and multi-scale pathways to effectively repurpose pretrained ViTs for dense outputs while retaining their generalization benefits. Mask2Former \cite{cheng2022masked} further unifies segmentation tasks through masked attention and set prediction, and has been adopted in pathology to better model heterogeneous structures and ambiguous boundaries. Together, these advances motivate integrating large pretrained representations with segmentation-oriented designs that are data-efficient, robust to domain shifts, and suitable for reliable TME quantification.

\section{Benchmark Design }
To systematically investigate to what extent token-level representations from pathology foundation models improve dense prediction, we introduce a standardized benchmarking framework for pathology segmentation.

As shown in Figure.~1, this benchmark spans the full evaluation pipeline, including diverse dense prediction datasets, a wide range of PFMs, multiple adaptation strategies, and unified evaluation metrics. Its purpose is to provide a comprehensive and reproducible assessment of the representation quality of PFMs, as well as to characterize the practical limits of their transferability to dense prediction tasks.
\begin{figure*}[t]
    \centering
    \includegraphics[width=\textwidth]{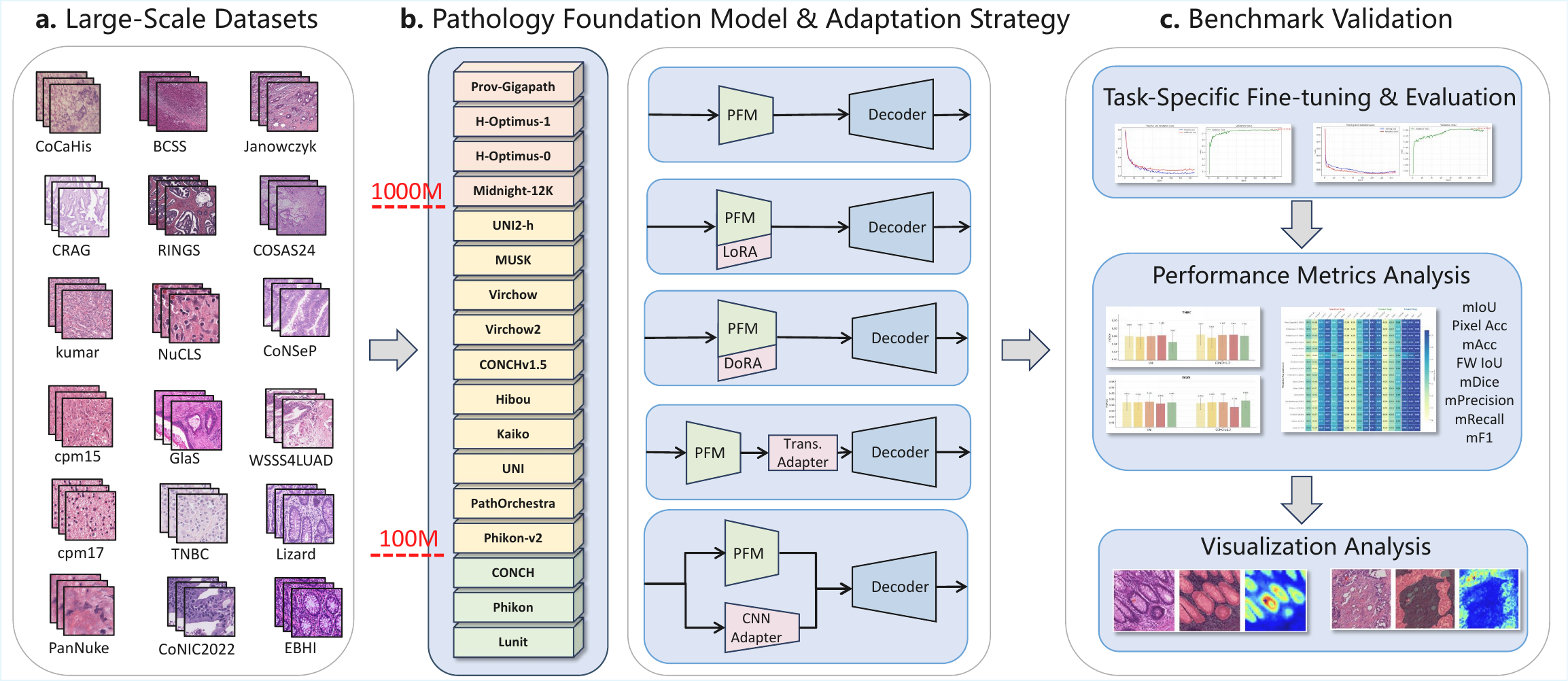}
    \caption{\textbf{Overview of PFM-DenseBench: A unified benchmark for evaluating Pathology Foundation Models on dense prediction.} The framework comprises three stages: \textbf{(1) Dataset Curation:} 18 public datasets covering nuclei-, gland-, and tissue-level segmentation across multiple organs. \textbf{(2) Model and Strategy Evaluation:} 17 vision-only and vision-language PFMs evaluated under five adaptation strategies, including LoRA/DoRA and CNN/Transformer adapters. \textbf{(3) Benchmark Validation:} A standardized protocol with task-specific fine-tuning, multi-metric evaluation, and qualitative analysis to characterize transfer effectiveness and scaling behavior.}

    \label{fig:twocolumn}
\end{figure*}

\subsection{Datasets}

% We evaluate the generalization of PFM representations on a standardized suite of 18 segmentation datasets from the PathSeg benchmark, designed to probe performance along three key axes:
We evaluate PFM representation generalization on 18 segmentation datasets, probing three key axes:

\begin{itemize}

    \item \textbf{Diverse Organ Systems.} The benchmark spans multiple anatomical sites, including Breast (BCSS \cite{amgad2019structured}, TNBC \cite{naylor2018segmentation}, NuCLS \cite{amgad2022nucls}), Colon (CONIC2022 \cite{graham2024conic}, GlaS \cite{sirinukunwattana2017gland}, Lizard \cite{graham2021lizard}), Lung (WSSS4LUAD \cite{han2022wsss4luad}), Prostate (RINGS \cite{salvi2021hybrid}), as well as multi-organ datasets covering Kidney, Liver, Stomach, and Pancreas (Kumar \cite{graham2020dense}, COSAS24, PanNuke \cite{gamper2019pannuke}).

    \item \textbf{Varied Cancer and Tissue Types.} The datasets encompass diverse pathological conditions, including adenocarcinoma (COSAS24, CRAG \cite{graham2019mild}), aggressive subtypes such as TNBC, and heterogeneous tissue components ranging from normal and malignant epithelium to inflammatory cells (e.g., CONIC2022 \cite{graham2024conic}) and tumor-associated stroma (e.g., BCSS \cite{amgad2019structured}).

    \item \textbf{Multi-Scale Annotation Granularity.} Tasks are grouped into nuclei-level segmentation (e.g., CONIC2022 \cite{graham2024conic}, PanNuke \cite{gamper2019pannuke}, CPM17 \cite{vu2019methods}), gland/structure-level segmentation (GlaS \cite{sirinukunwattana2017gland}, CRAG \cite{graham2019mild}, RINGS \cite{salvi2021hybrid}), and region-level tissue segmentation (BCSS \cite{amgad2019structured}, WSSS4LUAD \cite{han2022wsss4luad}), enabling evaluation from cellular to tissue scales.

\end{itemize}
Detailed specifications and the taxonomy of all 18 datasets are provided in Appendix A.

\subsection{Pathology Foundation Models}

We evaluate 17 state-of-the-art pathology foundation models (PFMs), grouped by pre-training paradigm into \emph{Vision-only} and \emph{Vision-Language} models.

\textbf{Vision-only PFMs.} These models are pre-trained on large-scale histopathology images to learn visual representations. Early work such as Phikon demonstrated the effectiveness of ViTs using iBOT. UNI later established a widely adopted baseline by pre-training DINOv2 on over 100k WSIs, a strategy followed by Hibou, H-Optimus-0, and H-Optimus-1 at million-scale. Subsequent models emphasize data and model scaling: Phikon v2 and UNI v2 incorporate IHC data, while Virchow adopts ViT-H/14 trained on millions of slides. Virchow2 further scales to over 3M slides with a 1.9B-parameter ViT-G/14. Beyond patch-level pre-training, Prov-GigaPath extends representations to the whole-slide level. We additionally include diverse baselines such as PathOrchestra, Midnight12k, Lunit, and Kaiko.

\textbf{Vision-Language PFMs.} These models incorporate textual supervision to align visual features with pathological semantics. CONCH adopts the CoCa framework to pre-train on large-scale pathology image--text pairs, and CONCH~1.5 further scales both data and model capacity. MUSK follows a different paradigm, using BEiT3 for unified masked modeling across vision and language.

Detailed introductions to these pathology foundation models are provided in Appendix B.

\subsection{Fine-tuning Strategies}
\begin{figure*}[t]
    \centering
    \includegraphics[width=\textwidth]{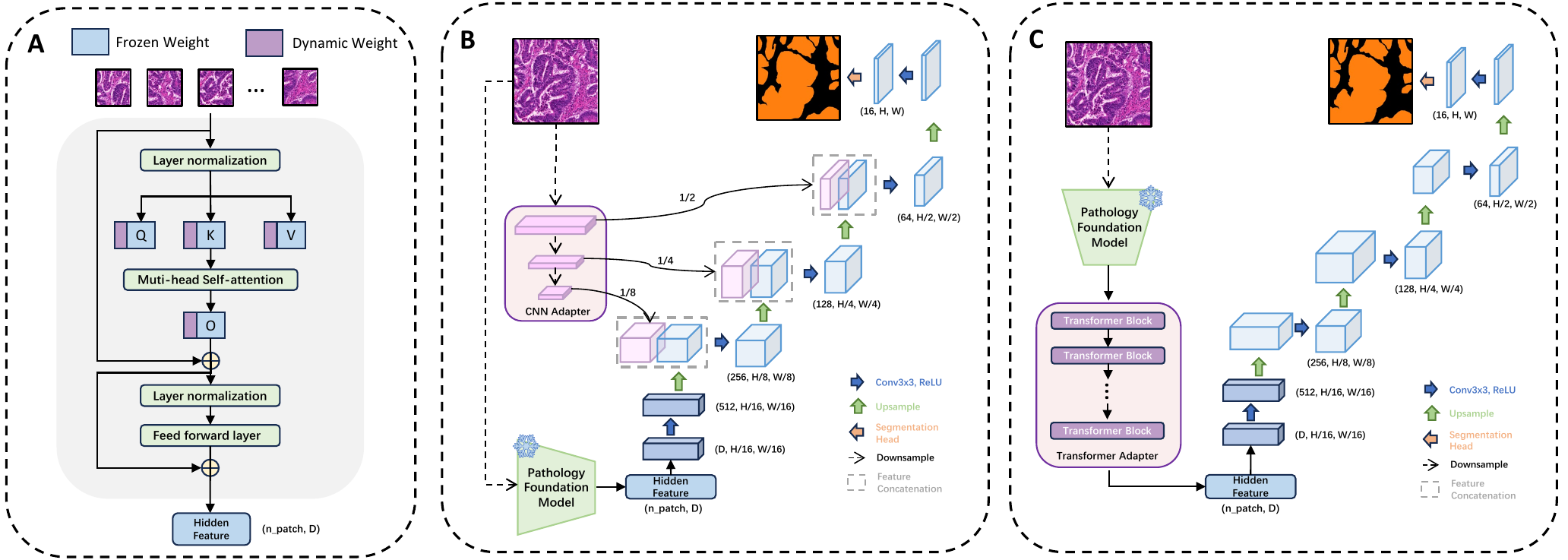}
    \caption{\textbf{Schematic illustration of parameter-efficient adaptation strategies for dense prediction with frozen Pathology Foundation Models.} \textbf{(A) Low-Rank Adaptation (LoRA/DoRA):} Trainable low-rank dynamic weights (purple) are injected into the Query (Q), Key (K), Value (V), and Output (O) projections of the frozen self-attention layers, decoupling optimization from the massive encoder parameters. \textbf{(B) CNN Adapter:} A parallel ResNetV2-style CNN branch extracts multi-scale local features alongside the frozen PFM. These features are injected into the decoder via skip connections to recover fine-grained spatial details and enhance boundary delineation. \textbf{(C) Transformer Adapter:} A sequential adaptation module that appends trainable Transformer blocks to the frozen encoder. This strategy processes the full token sequence to refine global semantic representations specifically for the downstream segmentation task.  }
    \label{fig:twocolumn}
\end{figure*}
\subsubsection{Frozen}
The PFM encoder is fully frozen, and only the decoder and segmentation head are trained. This setting isolates the contribution of pretrained token representations with minimal trainable parameters.

\subsubsection{LoRA}
As shown in Figure.~2(A), LoRA \cite{hu2022lora} freezes the encoder and injects trainable low-rank adapters into the self-attention layers. We apply LoRA to the QKV and output projections, optimizing only the low-rank parameters.

\subsubsection{DoRA}
As shown in Figure.~2(A), DoRA \cite{liu2024dora} augments low-rank adaptation with a learnable magnitude parameter to decouple weight direction and scale. The encoder remains frozen, and only the low-rank factors and magnitude parameter are trained.

\subsubsection{CNN Adapter}
As shown in Figure.~2(B), the CNN Adapter follows the TransUNet paradigm \cite{chen2024transunet} by fusing CNN spatial priors with Transformer tokens. We freeze the PFM encoder, add a parallel ResNetV2-style CNN branch to extract multi-scale features, and inject them into a skip-aware decoder. Only the CNN adapter, decoder, and segmentation head are trained.

\subsubsection{Transformer Adapter}
As shown in Figure.~2(C), the Transformer Adapter appends a lightweight stack of Transformer blocks to a frozen PFM encoder for parameter-efficient task specialization \cite{jose2025dinov2}. The full token sequence is processed by the appended blocks, after which only patch tokens are forwarded to the decoder. We train only the appended Transformer blocks, decoder, and segmentation head.

\begin{figure*}[t]
    \centering
    \includegraphics[width=\textwidth]{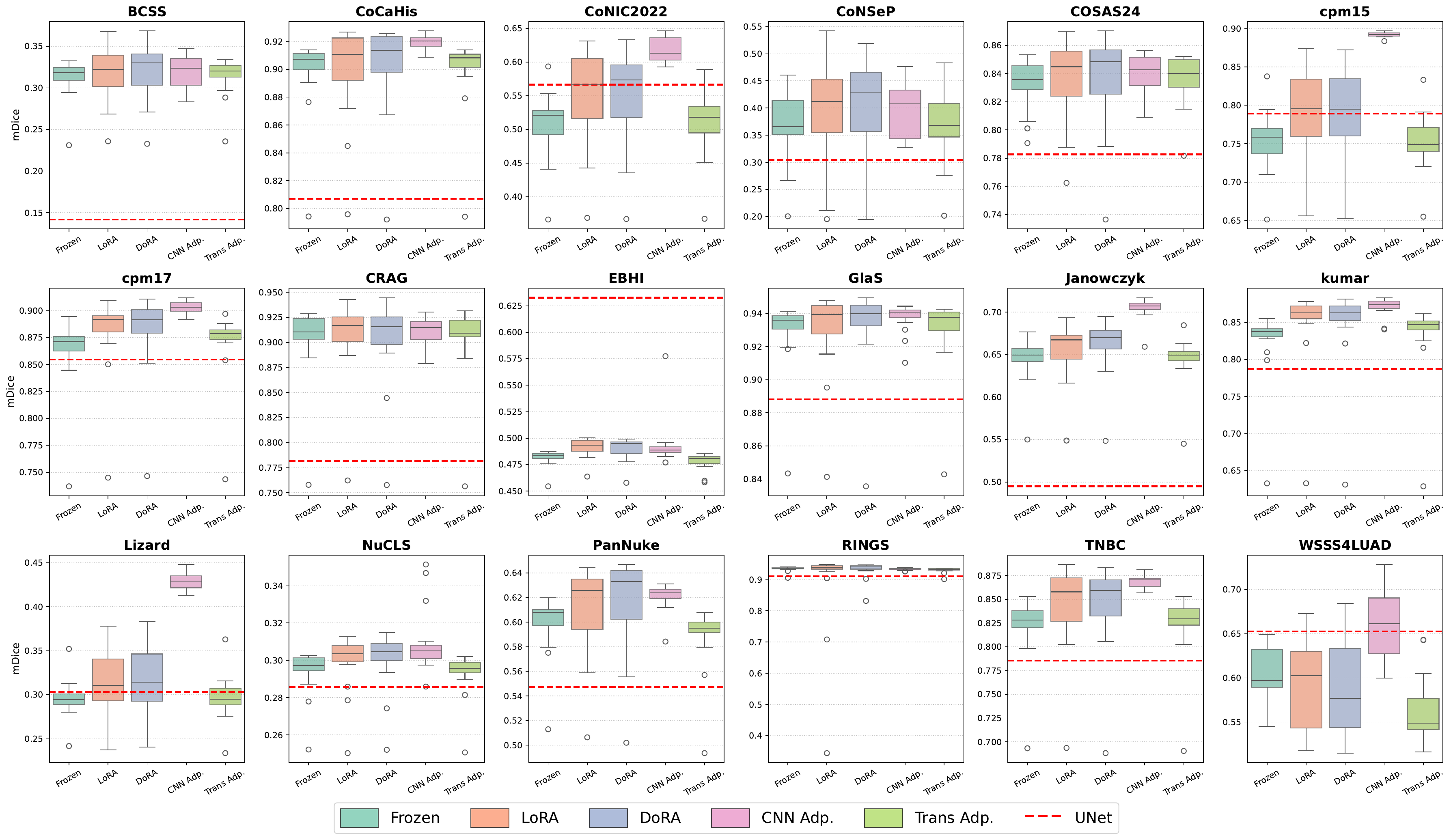}
    \caption{Benchmarking Pathology Foundation Models for Dense Prediction: Segmentation Performance Across Datasets and  Fine-tuning Strategies. Each box aggregates the evaluation results across 17 PFMs for a given fine-tuning strategie.   }
    \label{fig:twocolumn}
\end{figure*}

\subsection{Statistical analysis}
We quantify uncertainty in test-set performance using a nonparametric bootstrap over test images. Specifically, given \(N\) test samples, we generate \(B = 1000\) bootstrap replicates by sampling \(N\) samples with replacement from the test set. For each replicate, we compute the evaluation metric(s) on the resampled set, yielding an empirical sampling distribution per metric. We report the bootstrap mean and construct 95\% confidence intervals using the percentile method (2.5th and 97.5th percentiles of the \(B\) bootstrap values). The procedure is made reproducible by fixing the bootstrap random seed. 

\section{Experiments: Unlocking the Scaling Limits of Pathology Dense Prediction  }
Prior work on pathology foundation models has largely emphasized image-level recognition, leaving dense prediction relatively underexplored. In this section, we move beyond simple leaderboards to analyze the behavior of PFMs under dense constraints. Our experiments reveal a consistent narrative: while PFMs provide strong generalized features, naive scaling—whether in model size or adaptation parameters—yields diminishing returns. Instead, performance in dense prediction is driven by the alignment between token granularity, input resolution, and architectural inductive biases.

\subsection{Transferability of Foundation Models (RQ1 \& RQ2) }
We first establish whether the pre-trained representations of PFMs confer a tangible advantage over training from scratch. We evaluate 17 PFMs against a robust supervised UNet baseline across 18 datasets spanning nuclei, gland, and tissue segmentation tasks.

\textbf{Adaptation strategies are decisive:} While the "Frozen" setting (training only the decoder) offers a strong starting point, unlocking the full potential of PFMs requires targeted adaptation.

\textbf{The superiority of Convolutional biases:} Notably, the CNN Adapter frequently emerges as the top-performing strategy, particularly in challenging, fine-grained tasks such as CoNIC2022 and CPM15. This suggests that injecting local inductive biases via convolutional layers effectively complements the global semantic context captured by Vision Transformers.

\subsection{The Breakdown of Scaling Laws (RQ3)   }
A central tenet of recent deep learning research is that increasing model capacity—either through parameter count or adaptation rank—yields predictable performance gains. Our empirical analysis challenges this assumption in the context of pathology dense prediction.
\subsubsection{Model Scale is Not Predictive   }
We examined the relationship between model size (ranging from 21.7M to over 1B parameters) and segmentation performance under a frozen protocol. As shown in Figure.~4, segmentation performance does not exhibit a monotonic relationship with model size. Smaller or mid-sized models often match or outperform substantially larger ones.

\begin{figure}[htbp]
\vskip 0.2in
\begin{center}
\centerline{\includegraphics[width=\columnwidth, keepaspectratio]{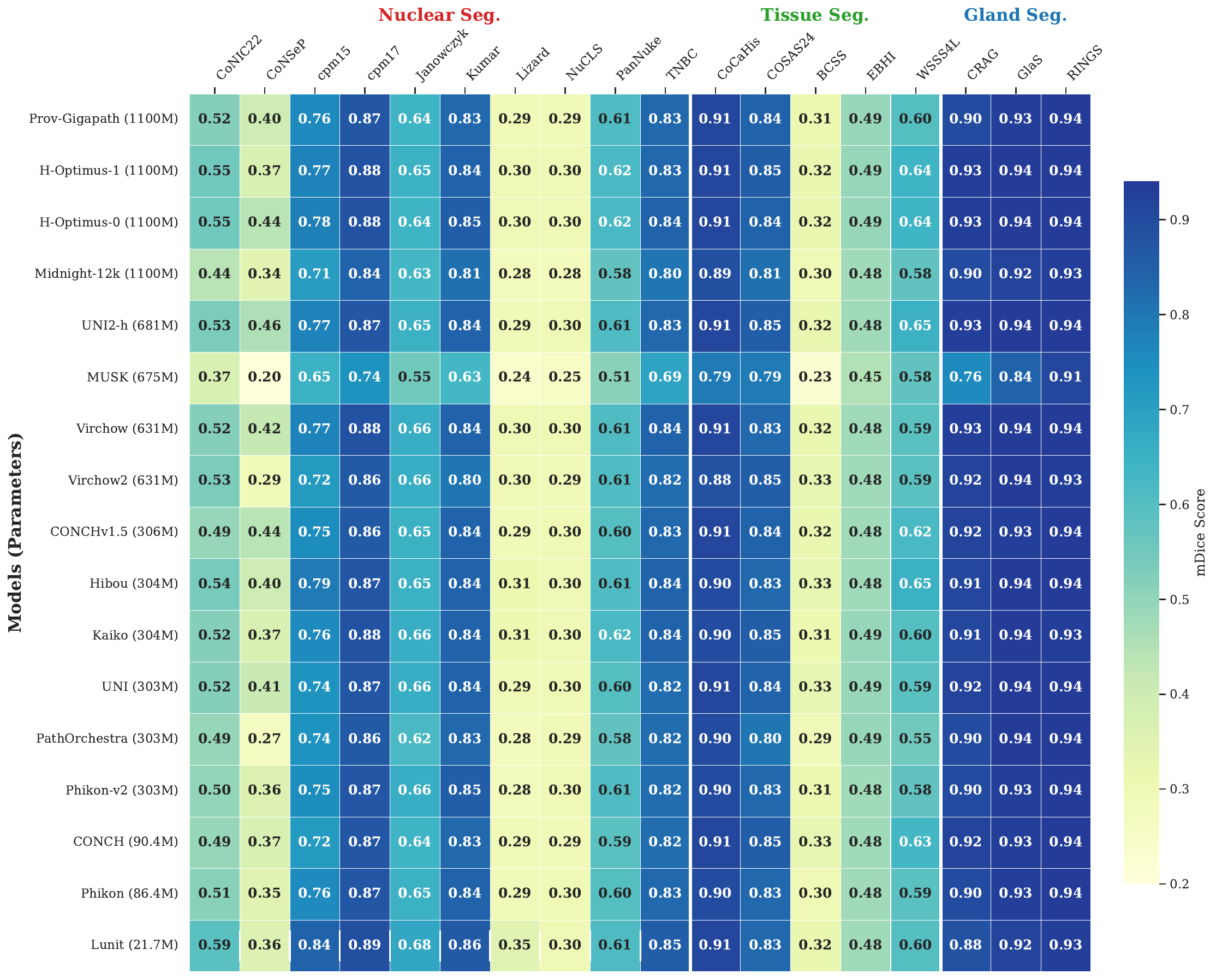}}
\caption{Scaling behavior of pathology foundation models under frozen method for dense prediction.   }
\label{icml-historical}
\end{center}
\vskip -0.2in
\end{figure}

\textbf{Interpretation:} This non-monotonicity suggests that pre-training objectives emphasizing global invariance (e.g., for slide-level classification) may induce excessive abstraction in larger models, potentially discarding the high-frequency spatial details required for boundary delineation. Parameter count alone is thus an unreliable proxy for dense prediction utility.

\subsubsection{Adaptation Capacity Saturates Early}
To further probe scaling limits, we investigated the impact of the rank in Low-Rank Adaptation (LoRA). Standard intuition suggests that higher ranks (more trainable parameters) should allow for better task adaptation. However, our ablation study on TNBC, GlaS, and COSAS24 (Figure.~5) reveals that increasing LoRA rank does not yield consistent monotonic improvements.

\textbf{Observation:} Performance often saturates or exhibits mild fluctuations, with overlapping confidence intervals between Rank 16 and Rank 128.

\textbf{Implication:} This suggests that the bottleneck in adapting PFMs to segmentation is not the capacity of the adapter, but rather the intrinsic structure of the frozen representations. Moderate ranks provide a favorable accuracy-efficiency trade-off, while larger ranks offer diminishing returns.

\begin{table}[htpb]
\caption{Effect of input resolution under frozen transfer on TNBC using UNI (vision-only). We vary the input size while keeping the ViT patch size fixed, and report mDice/mIoU with 95\% bootstrap confidence intervals.}
\label{tab:tnbc_resolution_ablation}
\vskip 0.15in
\begin{center}
\begin{small}
\begin{sc}
\begin{tabular}{ccc}
\toprule
Input Size & mDice (95\% CI) & mIoU (95\% CI) \\
\midrule
\texttt{256}  & 0.7323 {\tiny [0.7043, 0.7630]} & 0.6195 {\tiny [0.5847, 0.6554]} \\
\texttt{512}  & 0.8229 {\tiny [0.7875, 0.8523]} & 0.7232 {\tiny [0.6760, 0.7633]} \\
\texttt{1024} & \textbf{0.8456} {\tiny [0.8143, 0.8719]} & \textbf{0.7527} {\tiny [0.7099, 0.7886]} \\
\texttt{2048} & 0.8312 {\tiny [0.8017, 0.8570]} & 0.7365 {\tiny [0.7006, 0.7690]} \\
\texttt{4096} & 0.8386 {\tiny [0.8182, 0.8532]} & 0.7447 {\tiny [0.7166, 0.7647]} \\
\bottomrule
\end{tabular}
\end{sc}
\end{small}
\end{center}
\vskip -0.1in
\end{table}

\subsection{Resolution and Granularity: The True Drivers of Performance (RQ4)   }
If scaling model size and adapter rank fails to boost performance, what factors are critical? Our results point to input resolution and the resulting token granularity as the primary determinants of success.

\textbf{The "Sweet Spot" of Input Resolution:} Vision Transformers process images as sequences of fixed-size patches. Consequently, changing the input resolution implicitly alters the physical tissue area covered by each token. We evaluated the UNI model on the TNBC dataset in resolutions ranging from $256^2$ to $4096^2$ (Table 1).

\textbf{Resolution-Patch Mismatch:} We observe a clear dependence on input resolution. Increasing the input size from 256 to 1024 yields substantial gains, with mDice peaking at \ $ 1024 \times 1024 \ $.

\textbf{Diminishing Returns at High Res:} Surprisingly, pushing resolution further to 2048 or 4096 does not improve performance and can even cause slight degradation.

\begin{figure}[htbp]
\vskip 0.2in
\begin{center}
\centerline{\includegraphics[width=\columnwidth, keepaspectratio]{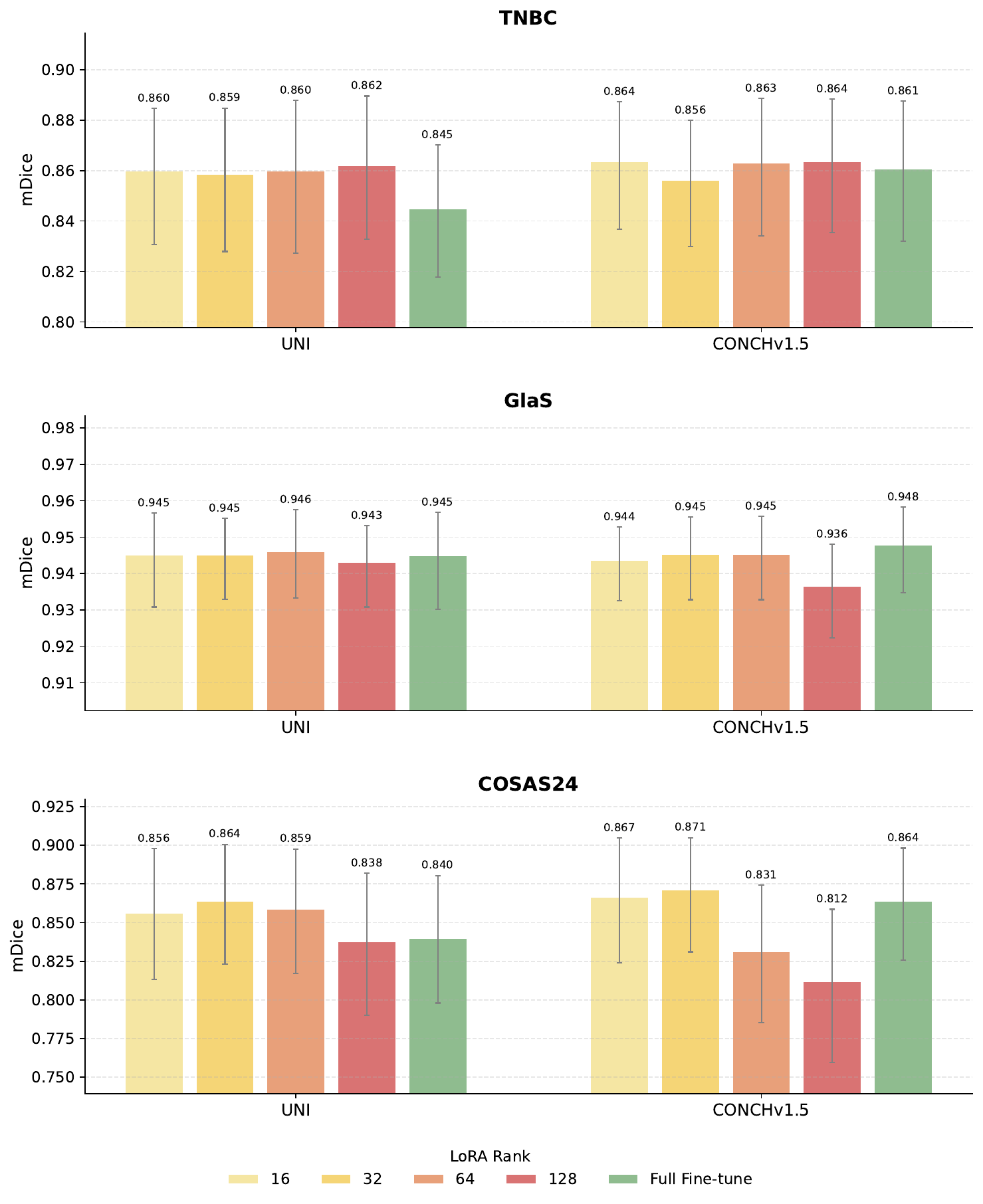}}
\caption{LoRA rank ablation across representative pathology segmentation regimes with 95\% bootstrap confidence intervals. We evaluate LoRA fine-tuning with varying ranks and full fine-tuning on three representative datasets spanning cell (TNBC), gland (GlaS), and tissue (COSAS24) segmentation.
}
\label{icml-historical}
\end{center}
\vskip -0.2in
\end{figure}

\begin{figure*}[!t]
    \centering
    \includegraphics[width=\textwidth]{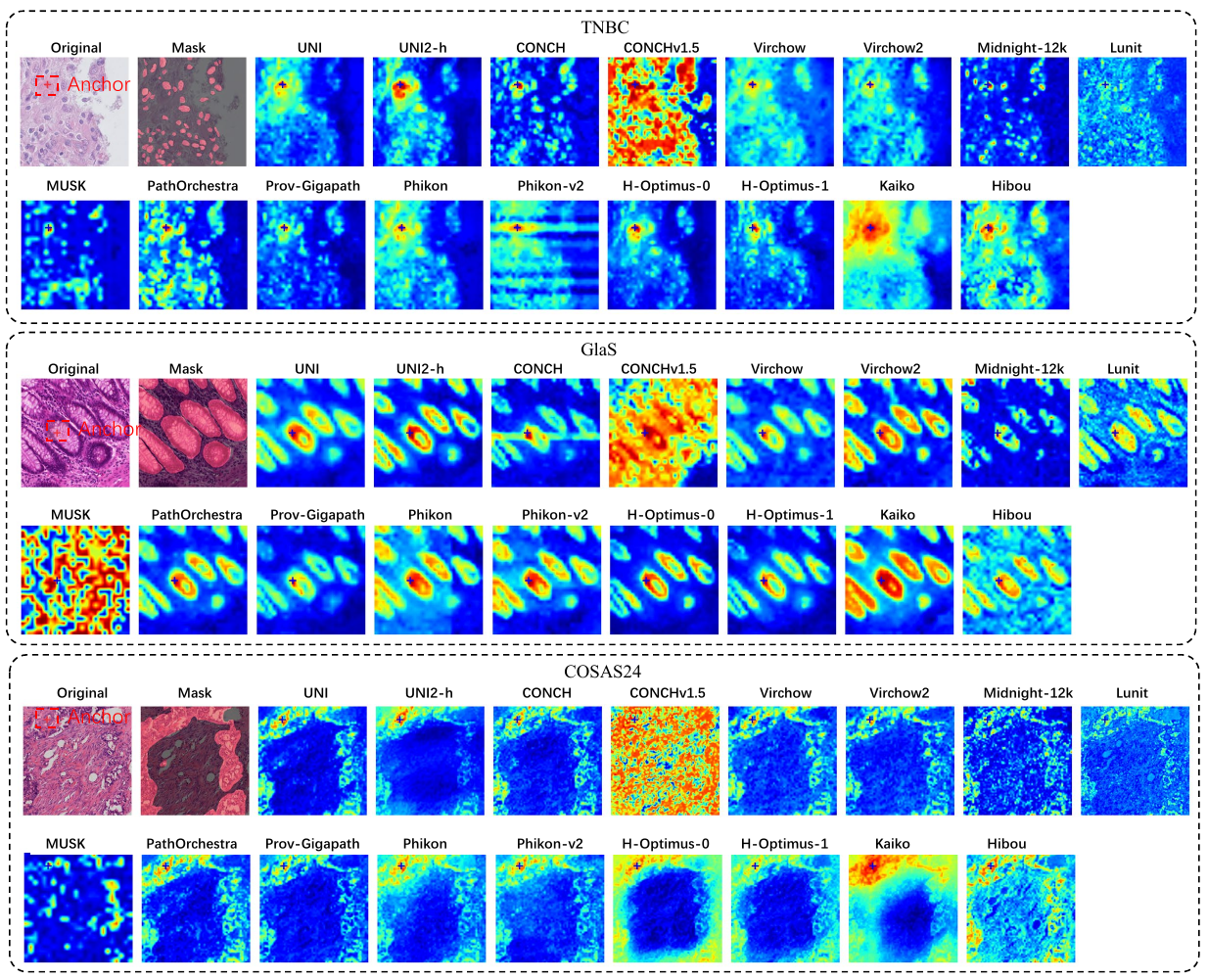}
    \caption{Cross-model comparison of last-layer patch-token similarity maps across pathology segmentation regimes.For each image, the red cross marks the query location; we compute a similarity heatmap by measuring cosine similarity between the last-layer patch token at the query and all other patch tokens (warmer colors indicate higher similarity). We show representative samples from TNBC (cell-level), GlaS (gland-level), and COSAS24 (tissue-level), comparing token-level spatial organization across pathology foundation models. }
    \label{fig:twocolumn}
\end{figure*}

\textbf{Conclusion:} At low resolutions, patch tokens aggregate too much distinct morphological content (e.g., multiple cells in one token), causing information loss. Conversely, at excessively high resolutions, the representation fragments into overly local sub-cellular patterns, losing the mid-range coherence necessary for segmentation.

\subsection{Mechanistic Interpretation via Token Similarity}

To explain why resolution impacts performance while LoRA rank does not, we visualize the spatial structure of the learned representations usinglast-layer patch-token similarity maps.

% \textbf{Why LoRA Scaling Fails:} Figure.~7 compares similarity maps across varying LoRA ranks (16 to 128). Strikingly, the similarity patterns remain largely unchanged: the spatial extent and contrast of high-affinity regions are stable. This qualitative invariance explains the quantitative saturation observed in Section 4.2: simply adding more trainable parameters to the adapter does not substantially reconfigure the underlying spatial organization of the frozen tokens.
\textbf{Why LoRA Scaling Fails:} Figure.~7 compares similarity maps across varying LoRA ranks (16 to 128). Strikingly, the similarity patterns remain largely unchanged: the spatial extent, topology, and contrast of high-affinity regions are stable, and the locations of peak responses exhibit minimal drift. This qualitative invariance provides a direct visual explanation for the quantitative saturation observed in Section~4.2. Intuitively, while higher-rank LoRA increases the adapter’s expressivity, it primarily induces small, low-frequency feature refinements rather than the kind of geometric reorganization needed for dense prediction. As a result, simply adding more trainable parameters to the adapter does not substantially reconfigure the underlying spatial organization of the frozen tokens, nor does it improve boundary delineation or fine-grained correspondence. In other words, LoRA rank scaling tends to amplify \emph{how much} the model can adjust within an existing representation manifold, but not \emph{where} spatial evidence is routed—leading to stable similarity maps and, consequently, performance saturation.

\begin{figure}[htbp]
\vskip 0.2in
\begin{center}
\centerline{\includegraphics[width=\columnwidth, keepaspectratio]{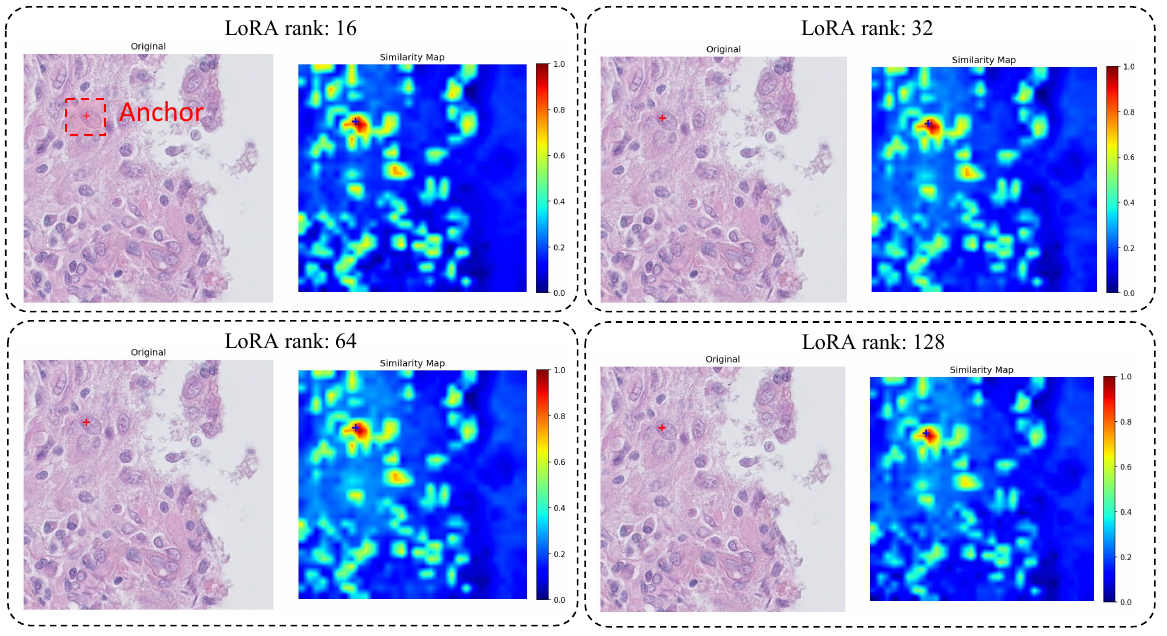}}
\caption{LoRA rank has limited impact on token-level spatial organization in frozen representations. Using the same query location (red cross), we visualize last-layer patch-token similarity maps for LoRA with different ranks (e.g., 16/32/64/128). Similarity patterns remain largely unchanged across ranks, supporting the rank-saturation behavior observed in Sec. 4.2.}
\label{icml-historical}
\end{center}
\vskip -0.2in
\end{figure}
\textbf{Why Resolution Matters:} In contrast, varying input resolution dramatically reshapes the representation (Figure.~8). As resolution increases from 256 to 1024, the similarity maps become progressively more structured, with regions corresponding to cellular morphology exhibiting stronger, sharper, and more coherent affinities. The boundaries between tissue components also become better defined, suggesting improved spatial correspondence and reduced mixing across distinct micro-structures. However, beyond 1024, the representation begins to fragment: affinities break into small scattered peaks, and tokens become overly localized, weakening long-range consistency and disrupting region-level continuity. This behavior aligns closely with the performance peak observed in Section~4.3, indicating a sweet spot where resolution provides sufficient morphological detail without sacrificing contextual integration. Overall, these results confirm that optimal dense prediction requires matching the effective patch granularity to the scale of the biological object of interest, balancing fine-grained detail with enough receptive field to maintain coherent tissue-level organization.

\begin{figure}[htbp]
\vskip 0.2in
\begin{center}
\centerline{\includegraphics[width=\columnwidth, keepaspectratio]{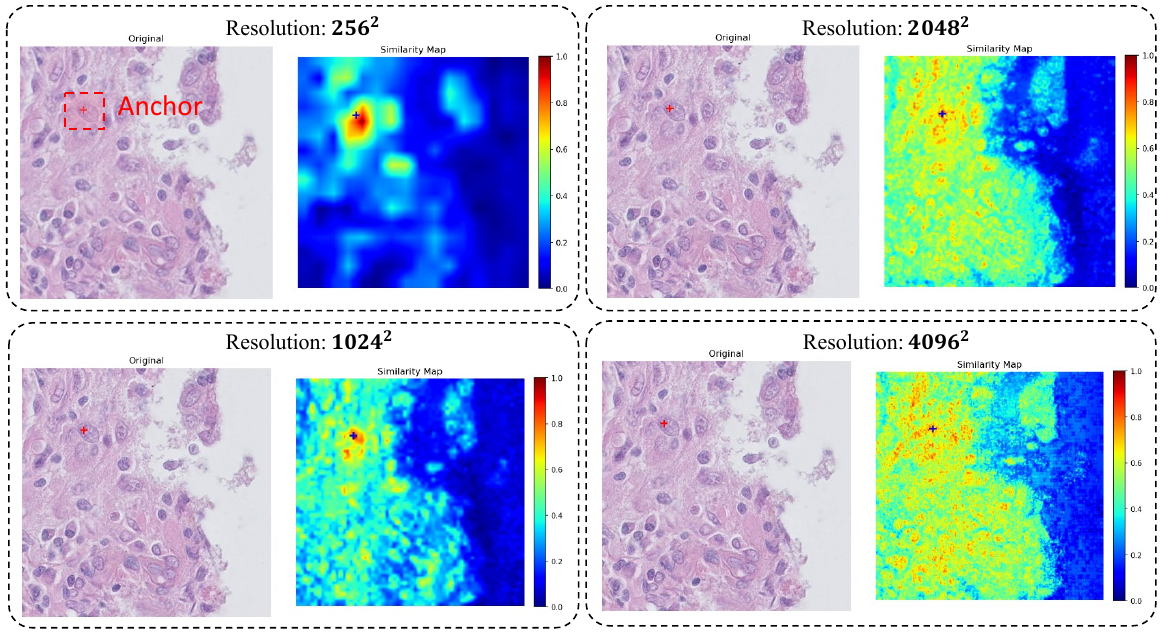}}
\caption{Input resolution reshapes patch-token similarity under a fixed patch size, revealing an optimal granularity regime.With the red cross denoting the query location, we compare similarity maps under varying input resolutions (256–4096) while keeping the ViT patch size fixed. Similarity structure strengthens from 256 to 1024, but becomes overly localized at higher resolutions, consistent with the resolution peak in Sec. 4.3. }
\label{icml-historical}
\end{center}
\vskip -0.2in
\end{figure}
In summary, our experiments demonstrate that transferring Pathology Foundation Models to dense prediction is not a problem of “more scale,” but one of \textbf{representation alignment}. Rather than relying on larger backbones or higher LoRA ranks, success depends on (i) choosing architectures that inject locality-preserving inductive biases (e.g., CNN adapters) and (ii) selecting an input resolution that preserves morphological integrity while maintaining coherent spatial context.

\section{Conclusions}
In this paper, we present \textbf{PFM-DenseBench}, a large-scale evaluation of 17 Pathology Foundation Models across 18 segmentation datasets. Our findings reveal that the transferability of PFMs to dense prediction is governed by representation alignment rather than naive scaling. Specifically, we demonstrate that increasing model size and adaptation rank yields diminishing returns, while performance is primarily driven by input resolution and convolutional inductive biases. Notably, this observation resonates with recent evidence from DINOv3 \cite{simeoni2025dinov3}, where a distillation-trained ConvNeXt \cite{liu2022convnet} backbone exhibits stronger dense prediction capability than substantially larger ViT models, underscoring the importance of locality-preserving architectural priors and offering a new direction for backbone design in pathology foundation model pretraining. These insights highlight a paradigm shift: to bridge the gap between emergent AI intelligence and precise clinical quantification, the field must transition from classification-centric backbones toward dense-representation-centered next-generation architectures. By releasing our unified benchmarking framework, we provide a roadmap for spatially precise foundation models that capture complex morphological hierarchies of human tissue. Moreover, we identify failure modes where global features overlook fine boundaries and rare micro-structures, motivating spatial diagnostics beyond aggregate scores.


% \section*{Software and Data}

% \section*{Acknowledgements}

\section*{Impact Statement}
This research provides a critical evaluation of the current landscape of Pathology Foundation Models (PFMs), offering the following contributions to the scientific community and clinical field:

\textbf{Comprehensive benchmarking and practical insights.} This work provides the computational pathology community with the first large-scale, systematic evaluation of how pathology foundation models (PFMs) perform in dense prediction tasks. By establishing a unified protocol across 18 datasets, we offer a definitive roadmap for model selection and adaptation, moving the field beyond image-level classification. Our findings on the interplay between token granularity, input resolution, and architectural inductive biases provide actionable guidance for researchers aiming to achieve pixel-level precision and reliable quantification of complex histological structures across heterogeneous datasets.

\textbf{Rethinking scaling laws and next-generation architectures.} Our study provides a critical reassessment of scaling laws in computational pathology, demonstrating that naive increases in parameter size or adaptation capacity often yield diminishing returns for dense tasks. This discovery challenges the prevailing "scale-at-all-costs" paradigm and shifts the focus toward the intrinsic spatial structure and integrity of learned features. By revealing these limitations, our work serves as a foundational blueprint for dense representation centered next-generation architectures that explicitly prioritize high-frequency spatial hierarchies. We believe this shift is essential for developing a new wave of AI tools capable of mastering the complex morphological landscape of human disease with unprecedented precision.

\nocite{langley00}

\bibliography{example_paper}
\bibliographystyle{icml2026}

%%%%%%%%%%%%%%%%%%%%%%%%%%%%%%%%%%%%%%%%%%%%%%%%%%%%%%%%%%%%%%%%%%%%%%%%%%%%%%%
%%%%%%%%%%%%%%%%%%%%%%%%%%%%%%%%%%%%%%%%%%%%%%%%%%%%%%%%%%%%%%%%%%%%%%%%%%%%%%%
% APPENDIX
%%%%%%%%%%%%%%%%%%%%%%%%%%%%%%%%%%%%%%%%%%%%%%%%%%%%%%%%%%%%%%%%%%%%%%%%%%%%%%%
%%%%%%%%%%%%%%%%%%%%%%%%%%%%%%%%%%%%%%%%%%%%%%%%%%%%%%%%%%%%%%%%%%%%%%%%%%%%%%%
\newpage
\appendix
\onecolumn
\section{Datasets }
\renewcommand{\thefigure}{A\arabic{figure}}
\renewcommand{\thetable}{A\arabic{table}}
\setcounter{figure}{0}
\setcounter{table}{0}
{Breast cancer semantic segmentation (BCSS):} Derived from The Cancer Genome Atlas (TCGA-BRCA) cohort, the BCSS dataset serves as a fundamental resource for tissue-level segmentation in breast pathology, specifically focusing on the aggressive Triple Negative Breast Cancer (TNBC) subtype. It contains over 20,000 annotations generated through the collaborative efforts of pathologists and medical students, dividing tissue regions into biologically distinct compartments: tumor, stroma, inflammatory infiltrates, and necrosis. This dataset is critical for training models to quantify the Tumor-Stroma Ratio (TSR), a known independent prognostic factor, and provides the macro-architectural context necessary for distinguishing tumor epithelium from associated stroma. Availability:  \href{https://bcsegmentation.grand-challenge.org/}{https://github.com/PathologyDataScience/BCSS}

\textbf{Colon cancer histopathological dataset (CoCaHis):} The CoCaHis dataset specifically addresses the domain shift challenge presented by intraoperative frozen sections, which differ significantly from standard Formalin-Fixed Paraffin-Embedded (FFPE) slides due to rapid preparation artifacts such as ice crystals and variable staining. It comprises 82 hematoxylin and eosin (H\&E) stained images from 19 patients diagnosed with metastatic colon cancer in the liver, accompanied by pixel-wise ground truth maps annotated by pathologists. This dataset is essential for validating the robustness of segmentation models in clinical intraoperative settings, requiring algorithms to ignore cryosection-specific noise while accurately identifying metastatic adenocarcinoma regions. Availability:  \href{https://www.researchgate.net/publication/348604843_A_dataset_and_a_methodology_for_intraoperative_computer-aided_diagnosis_of_a_metastatic_colon_cancer_in_a_liver}{https://cocahis.irb.hr/}

\textbf{Colon nuclei identification and counting (CONIC2022):} Developed for the CoNIC 2022 Challenge, this dataset represents one of the largest publicly available resources for nuclei-level analysis in computational pathology, containing approximately 500,000 labeled nuclei derived from the Lizard dataset. It focuses on the detailed mapping of the colonic tumor microenvironment by categorizing nuclei into six specific classes: epithelial, lymphocyte, plasma, eosinophil, neutrophil, and connective tissue. By necessitating the differentiation of inflammatory subtypes (e.g., distinguishing plasma cells from lymphocytes), the dataset facilitates the development of automated immune profiling tools relevant for predicting immunotherapy responses and patient survival. Availability:  \href{https://conic-challenge.grand-challenge.org/}{https://conic-challenge.grand-challenge.org/}

\textbf{Colorectal nuclear segmentation and phenotypes (CoNSeP):} CoNSeP is a high-fidelity dataset consisting of 41 H\&E stained image tiles ($1,000 \times 1,000$ pixels at $40\times$ magnification) extracted from 16 colorectal adenocarcinoma Whole Slide Images (WSIs). It is distinguished by its precise pixel-level annotations that separate nuclei into distinct classes, including malignant epithelial, normal epithelial, inflammatory, muscle, fibroblast, and endothelial cells. The dataset is particularly challenging due to the high density of overlapping nuclei and is frequently used to benchmark instance segmentation and classification models that must rely on texture and morphology rather than mere spatial detection. Availability:  \href{https://opendatalab.com/OpenDataLab/CoNSeP/tree/main}{https://opendatalab.com/OpenDataLab/CoNSeP/tree/main}

\textbf{Cross-organ and cross-scanner adenocarcinoma segmentation (COSAS24):} Introduced for the MICCAI 2024 challenge, COSAS24 is designed to benchmark domain generalization capabilities by incorporating data spanning multiple organs (gastric, colorectal, and pancreatic adenocarcinoma) and multiple digital scanners (TEKSQRAY, KFBIO, 3DHISTECH). It provides a rigorous testbed for evaluating whether segmentation algorithms can learn the intrinsic features of adenocarcinoma—such as glandular atypia and loss of polarity—independent of organ-specific background textures or scanner-induced color variations. This dataset is pivotal for developing foundation models capable of deploying reliably across diverse clinical centers. Availability:  \href{https://cosas.grand-challenge.org/}{https://cosas.grand-challenge.org/}

\textbf{Computational precision medicine 2015 (CPM15):} Originating from the MICCAI 2015 challenge, CPM15 is a foundational dataset for nuclear segmentation in brain tumors, containing images of Lower Grade Glioma (LGG) and Glioblastoma Multiforme (GBM) at $40\times$ magnification. While smaller in scale compared to modern benchmarks, it poses significant challenges due to the high pleomorphism of tumor nuclei and the overlapping boundaries characteristic of glial and neuronal cells. It remains a key resource for ensuring that general pathology models can handle non-epithelial tissue types and the unique morphological presentation of central nervous system malignancies. Availability:  \href{https://drive.google.com/drive/folders/1l55cv3DuY-f7-JotDN7N5nbNnjbLWchK}{https://drive.google.com/drive/folders/1l55cv3DuY-f7-JotDN7N5nbNnjbLWchK}

\textbf{Computational precision medicine 2017 (CPM17):} CPM17 extends the scope of the earlier CPM15 benchmark by incorporating a multi-organ cohort to evaluate the generalization of nuclear segmentation algorithms. It includes 32 training and 32 testing image tiles derived from TCGA samples of Head \& Neck, Lung, and Brain tissues, requiring models to handle significant variations in nuclear appearance and staining protocols across different anatomical sites. This dataset is widely used to validate the ability of deep learning models to abstract the concept of a "nucleus" beyond tissue-specific contexts. Availability:  \href{https://drive.google.com/drive/folders/1l55cv3DuY-f7-JotDN7N5nbNnjbLWchK}{https://drive.google.com/drive/folders/1l55cv3DuY-f7-JotDN7N5nbNnjbLWchK}

\textbf{Colorectal adenocarcinoma gland (CRAG):} The CRAG dataset focuses on the instance segmentation of glandular structures in colorectal adenocarcinoma, comprising 213 H\&E image tiles at $20\times$  magnification with varying cancer grades. It addresses the complexity of segmenting malignant glands that exhibit fused, cribriform, or deformed morphologies, which are critical features for histological grading. Success on this dataset demonstrates a model's ability to understand topological structures—specifically the epithelial-lumen organization—rather than relying solely on local textural cues. Availability:  \href{https://warwick.ac.uk/fac/cross\_fac/tia/data/mildnet/}{https://warwick.ac.uk/fac/cross\_fac/tia/data/mildnet/}

\textbf{Enteroscope biopsy histopathological image (EBHI):} EBHI-Seg is a comprehensive dataset designed to cover the entire progression spectrum of colorectal lesions, containing 4,456 histopathological images encompassing six distinct differentiation stages: normal, polyp, low-grade intraepithelial neoplasia (IN), high-grade IN, serrated adenoma, and adenocarcinoma. Unlike binary benign/malignant datasets, EBHI enables the training of fine-grained diagnostic models capable of identifying precancerous lesions, which is clinically vital for determining the urgency of surgical intervention and understanding the continuous biological spectrum of tumorigenesis. Availability:  \href{https://www.kaggle.com/datasets/alibabaei78/ebhi-seg}{https://www.kaggle.com/datasets/alibabaei78/ebhi-seg}

\textbf{Gland segmentation challenge (GlaS):} As the inaugural benchmark for gland segmentation from the MICCAI 2015 challenge, GlaS consists of 165 H\&E stained images from 16 colorectal carcinoma sections, covering both benign and malignant glandular structures. The primary technical challenge presented by this dataset is the separation of closely clustered glands, a requisite for accurate morphological assessment and tumor grading. It remains a standard benchmark for evaluating the efficacy of semantic segmentation models in distinguishing glandular architecture from the surrounding stroma. Availability:  \href{https://datasetninja.com/gland-segmentation#download}{https://datasetninja.com/gland-segmentation}

\textbf{Deep learning for digital pathology (Janowczyk):} Curated by Andrew Janowczyk, this series of datasets was originally developed for a comprehensive tutorial on deep learning in digital pathology and encompasses diverse primitives including nuclei, epithelium, tubules, mitosis, and lymphoma subtypes. It serves as a standardized, "textbook" quality benchmark that allows models to learn canonical pathological features across different staining and tissue contexts. The dataset is instrumental for initial model training and validation of fundamental segmentation capabilities before scaling to more noisy or complex clinical data. Availability:  \href{https://andrewjanowczyk.com/use-case-1-nuclei-segmentation/}{https://andrewjanowczyk.com/use-case-1-nuclei-segmentation/}

\textbf{Monuseg (Kumar):} The MoNuSeg (Multi-Organ Nucleus Segmentation) dataset, often referred to as the Kumar dataset, was the first to explicitly emphasize multi-organ generalization for nuclear segmentation. It contains 30 image tiles ($1,000 \times 1,000$ pixels) captured at $40\times$ magnification from TCGA, covering seven distinct organs: Breast, Kidney, Liver, Prostate, Bladder, Colon, and Stomach. The diversity of tissue backgrounds—from liver sinusoids to renal tubules—makes this dataset the primary benchmark for assessing a model's ability to segment nuclei robustly, independent of the underlying organ architecture. Availability:  \href{https://drive.google.com/drive/folders/1bI3RyshWej9c4YoRW-\_q7lh7FOFDFUrJ}{https://drive.google.com/drive/folders/1bI3RyshWej9c4YoRW-\_q7lh7FOFDFUrJ}

\textbf{Lizard (Lizard):} Lizard is a massive-scale dataset tailored for data-hungry deep learning models, containing nearly half a million labeled nuclei from colorectal histology images at $20\times$ magnification. It aggregates and refines data from multiple sources to provide high-granularity class labels such as epithelial cells, connective tissue, lymphocytes, plasma cells, neutrophils, and eosinophils. This dataset ensures that models trained on it possess both high precision in segmentation and the ability to perform detailed cellular phenotyping within the colon tissue context. Availability:  \href{https://www.kaggle.com/datasets/aadimator/lizard-dataset}{https://www.kaggle.com/datasets/aadimator/lizard-dataset}

\textbf{Nucleus classification, localization, and segmentation (NuCLS):} NuCLS is a large-scale dataset focusing on breast cancer that utilizes a crowdsourcing approach to generate over 220,000 labeled nuclei from TCGA images. It places a specific emphasis on the assessment of Tumor-Infiltrating Lymphocytes (TILs) by providing annotations that distinguish between tumor, stromal, and TIL nuclei. This distinction is biologically significant for breast cancer prognosis, and the dataset trains models to differentiate small, dark, round lymphocytes from morphologically similar tumor nuclei in dense tissue regions. Availability:  \href{https://nucls.grand-challenge.org/}{https://nucls.grand-challenge.org/}

\textbf{Pan-cancer nuclei (PanNuke):} PanNuke is an open, pan-cancer histology dataset designed to maximize tissue coverage, containing semi-automatically generated instance segmentation masks for over 200,000 nuclei across 19 different tissue types. The nuclei are categorized into five classes: Neoplastic, Inflammatory, Connective, Dead, and Non-Neoplastic Epithelial. By covering rare tissue types often omitted in other datasets, PanNuke is essential for training "foundation models" that aim to achieve universal segmentation capabilities across the human body. Availability:  \href{https://warwick.ac.uk/fac/sci/dcs/research/tia/data/pannuke}{https://warwick.ac.uk/fac/sci/dcs/research/tia/data/pannuke}

\textbf{Rapid identification of glandular structures (RINGS):} The RINGS dataset addresses the specific challenges of prostate cancer diagnosis, containing 1,500 H\&E stained images with 18,851 annotated glands. It focuses on the segmentation of prostate glands which, unlike colonic glands, often exhibit severe degeneration and irregular, sieve-like (cribriform) structures in higher-grade cancers. This dataset enables models to learn the complex topological variations required for automated Gleason grading, moving beyond simple shape detection to the recognition of disrupted architectural patterns. Availability:  \href{https://data.mendeley.com/datasets/h8bdwrtnr5/1}{https://data.mendeley.com/datasets/h8bdwrtnr5/1}

\textbf{Triple negative breast cancer (TNBC):} This dataset comprises 50 images from 11 Triple Negative Breast Cancer patients and is specifically curated to test segmentation performance in highly heterogeneous and aggressive disease subtypes. The images differ from standard breast cancer datasets by featuring extremely high cellularity and pleomorphism, serving as a stress test for algorithms to separate touching and overlapping nuclei in dense tumor regions where boundary delineation is particularly difficult. Availability:  \href{https://peterjacknaylor.github.io/data/}{https://peterjacknaylor.github.io/data/}

\textbf{Weakly supervised semantic segmentation for lung adenocarcinoma (WSSS4LUAD):} WSSS4LUAD focuses on the tissue-level semantic segmentation of Lung Adenocarcinoma, differentiating between tumor epithelium, tumor-associated stroma, and normal tissue. While originally designed for weakly supervised learning research, the test set provides dense pixel-level annotations that are crucial for training models to distinguish pulmonary architectural patterns, such as separating malignant solid nests or acini from normal alveolar structures and the desmoplastic stromal reaction unique to lung malignancies. Availability:  \href{https://wsss4luad.grand-challenge.org/}{https://wsss4luad.grand-challenge.org/}

\begin{table}[htbp]
  \centering
  \caption{An overview of the selected pathology image segmentation datasets in our benchmark. The table reports the dataset name, magnification, anatomical region, and input resize for ViT-14/ViT-16. (Note that all datasets were run at a fixed resolution of 384 on the MUSK model.) }
  \label{tab:dataset_overview}
  \begin{tabular}{lccc}
    \toprule
    \textbf{Dataset}& \textbf{Magnification}& \textbf{Anatomical region}& \textbf{Resize (ViT-14/16)}\\
    \midrule
    BCSS \cite{amgad2019structured}& 40x & Breast & 952 / 944 \\
    CoCaHis \cite{sitnik2021dataset}& 40x & Liver & 994 / 992 \\
    CoNIC2022 \cite{graham2024conic}& 20x & Colon & 252 / 256 \\
    CoNSeP \cite{graham2019hover}& 40x & Colon & 994 / 992 \\
    COSAS24& - & Unspecified & 994 / 992 \\
    CPM15 \cite{vu2019methods} &  20x, 40x & Unspecified & 392 / 384 \\
    CPM17 \cite{vu2019methods} &  20x, 40x & Unspecified & 490 / 496 \\
    CRAG \cite{graham2019mild} & 20x & Colon & 994 / 992 \\
    EBHI \cite{shi2023ebhi} & 40x & Colon & 224 / 224 \\
    GlaS \cite{sirinukunwattana2017gland} & 20x & Colon & 420 / 416 \\
    Janowczyk \cite{janowczyk2016deep} & 40x & Breast & 994 / 992 \\
    Kumar \cite{graham2020dense} & 40x & Unspecified & 994 / 992 \\
    Lizard \cite{graham2021lizard} & 20x & Unspecified & 336 / 336 \\
    NuCLS \cite{amgad2022nucls} & 40x & Breast & 252 / 256 \\
    PanNuke \cite{gamper2019pannuke} & 40x & Unspecified & 252 / 256 \\
    RINGS \cite{salvi2021hybrid} & 100x & Prostate & 994 / 992 \\
    TNBC \cite{naylor2018segmentation} & 40x & Breast & 504 / 512 \\
    WSSS4LUAD \cite{han2022wsss4luad} & 10x & Lung & 196 / 192 \\
    \bottomrule
  \end{tabular}
\end{table}

\section{Pathology Foundation Model }
\renewcommand{\thefigure}{B\arabic{figure}}
\renewcommand{\thetable}{B\arabic{table}}
\setcounter{figure}{0}
\setcounter{table}{0}
\textbf{PathOrchestra \cite{yan2025pathorchestra}}: As a comprehensive foundation model released in March 2026, PathOrchestra is designed to handle over 100 diverse clinical-grade tasks, distinguishing itself by its training on a massive dataset comprising 300,000 pathological slides from 20 different tissue and organ types across multiple centers. The model leverages self-supervised learning to establish a versatile baseline for tasks ranging from digital slide preprocessing and pan-cancer classification to lesion identification and structured report generation, aiming to orchestrate a unified approach to computational pathology. [\href{https://huggingface.co/AI4Pathology/PathOrchestra}{https://huggingface.co/AI4Pathology/PathOrchestra}]

\textbf{UNI \cite{chen2024towards}}: UNI is a groundbreaking vision-only foundation model that utilizes the DINOv2 framework and a ViT-L/16 architecture, pre-trained on a curated dataset of over 100 million patches extracted from 100,426 whole-slide images (WSIs) across 20 diverse tissue types. Developed using high-quality data from Mass General Brigham, it addresses the critical issue of data contamination by establishing a robust evaluation benchmark across 34 representative pathological tasks, setting a new standard for general-purpose pathology encoding. [\href{https://huggingface.co/MahmoodLab/UNI}{https://huggingface.co/MahmoodLab/UNI}]

\textbf{UNI2-h \cite{chen2024towards}}: Building upon the success of its predecessor, UNI2-h (referred to as UNI2 in the provided table) scales the architecture to a ViT-H/14 with 631 million parameters and expands the pre-training data to 350,000 whole-slide images, yielding approximately 200 million patches. This model incorporates both H\&E and IHC staining data to enhance its versatility and is trained using the DINOv2 strategy, further pushing the boundaries of performance in large-scale computational pathology tasks. [\href{https://huggingface.co/MahmoodLab/UNI2-h}{https://huggingface.co/MahmoodLab/UNI2-h}]

\textbf{CONCH \cite{lu2024visual}}: CONCH is a seminal vision-language foundation model that leverages the CoCa framework to integrate visual and textual modalities, pre-trained on a substantial collection of 1.17 million image-text pairs sourced from high-quality PubMed articles, educational materials, and proprietary in-house datasets. Using a ViT-B/16 vision encoder and a text encoder, it excels in tasks requiring cross-modal understanding such as image captioning and text-to-image retrieval, demonstrating the efficacy of learning from diverse medical literature. [\href{https://huggingface.co/MahmoodLab/CONCH}{https://huggingface.co/MahmoodLab/CONCH}]

\textbf{CONCHv1.5 \cite{ding2025multimodal}}: As an advanced iteration of the CONCH model, CONCHv1.5 scales up the architecture to a ViT-L/16 with 306 million parameters and is pre-trained on a larger corpus of 1.26 million image-text pairs. This version continues to utilize the CoCa framework to refine vision-language alignment, offering improved performance on downstream tasks and serving as the patch-level encoder for the multimodal TITAN model. [\href{https://huggingface.co/MahmoodLab/conchv1_5}{https://huggingface.co/MahmoodLab/conchv1\_5}]

\textbf{Virchow \cite{vorontsov2024foundation}}: Virchow represents a significant milestone as one of the first models to scale pre-training data to the million-slide level, utilizing 1.5 million whole-slide images to train a ViT-H/14 architecture with 631 million parameters via the DINOv2 algorithm. By leveraging a massive proprietary dataset from Memorial Sloan Kettering Cancer Center, it aims to capture a wide breadth of pancancer morphological diversity, establishing a strong baseline for large-scale vision-only pathology models. [\href{https://huggingface.co/paige-ai/Virchow}{https://huggingface.co/paige-ai/Virchow}]

\textbf{Virchow2 \cite{zimmermann2024virchow2}}: Extending the capabilities of the original Virchow, Virchow2 pushes the data scale further to over 3.1 million whole-slide images, resulting in the largest vision-only model in this collection with a ViT-H/14 architecture (noting the table lists ViT-H/14 while the text mentions ViT-G/14, the image confirms ViT-H/14 with 631M parameters for Virchow2 as well). Trained with DINOv2 on in-house data including both H\&E and IHC stains, it demonstrates superior generalization and robustness across a wide array of clinical benchmarks. [\href{https://huggingface.co/paige-ai/Virchow2}{https://huggingface.co/paige-ai/Virchow2}]

\textbf{Phikon \cite{Filiot2023ScalingSSLforHistoWithMIM}}: Phikon empirically validates the effectiveness of the Vision Transformer (ViT-B/16) architecture for pan-cancer representation learning, utilizing the iBOT framework on a dataset of 6,093 whole-slide images from The Cancer Genome Atlas (TCGA). With 86.4 million parameters, it serves as a robust baseline for publicly available foundation models, demonstrating that strong performance can be achieved with moderate-scale public datasets. [\href{https://huggingface.co/owkin/phikon}{https://huggingface.co/owkin/phikon}]

\textbf{Phikon-v2 \cite{filiot2024phikon}}: Phikon-v2 (referred to as Phikon2 in the table) significantly expands upon its predecessor by enlarging the pre-training dataset to 58,359 public whole-slide images and 456 million patches, scaling the architecture to a ViT-L/16 with 303 million parameters. Utilizing the DINOv2 training strategy, it integrates diverse public datasets to create a powerful, publicly accessible feature extractor that rivals models trained on private proprietary data. [\href{https://huggingface.co/owkin/phikon-v2}{https://huggingface.co/owkin/phikon-v2}]

\textbf{Prov-Gigapath \cite{xu2024whole}}: Prov-Gigapath is a pioneering model that broadens the scope of pathology modeling by first employing a whole-slide modeling approach, utilizing a massive dataset of 171,000 whole-slide images and 1.4 billion patches to train a ViT-G/14 architecture with 1.1 billion parameters. It combines DINOv2 with masked image modeling (MIM) to capture both local patch details and global slide-level context, setting state-of-the-art performance on tasks requiring long-context understanding. [\href{https://huggingface.co/prov-gigapath/prov-gigapath}{https://huggingface.co/prov-gigapath/prov-gigapath}]

\textbf{H-Optimus-0}: H-Optimus-0 is a large-scale vision foundation model employing the DINOv2 and iBOT frameworks, trained on a proprietary dataset of 500,000 whole-slide images. With a massive ViT-G/14 architecture comprising 1.1 billion parameters, it leverages the diversity of half a million slides to generate high-quality representations, demonstrating the benefits of scaling model size and data volume simultaneously. [\href{https://huggingface.co/bioptimus/H-optimus-0}{https://huggingface.co/bioptimus/H-optimus-0}]

\textbf{H-Optimus-1}: H-Optimus-1 represents the state-of-the-art in the Optimus series, scaling the pre-training data to over 1 million whole-slide images and 2.0 billion patches, resulting in a 1.1 billion parameter model based on the ViT-G/14 architecture. Released in February 2026, it utilizes advanced self-supervised learning techniques to capture intricate histological patterns across a vast diversity of tissue types and staining conditions. [\href{https://huggingface.co/bioptimus/H-optimus-1}{https://huggingface.co/bioptimus/H-optimus-1}]

\textbf{MUSK \cite{xiang2025vision}}: MUSK is a multimodal transformer that utilizes a unified masked modeling strategy based on the BEiT3 architecture to integrate vision and language, pre-trained on 33,000 whole-slide images and over 1 million image-text pairs. With 675 million parameters, it effectively aligns pathological images with textual descriptions from PubMed, TCGA, and other sources, enabling advanced capabilities in visual question answering and cross-modal retrieval. [\href{https://huggingface.co/xiangjx/musk}{https://huggingface.co/xiangjx/musk}]

\textbf{Midnight-12k \cite{KDK2025}}: Midnight-12k is an efficient pathology foundation model that challenges the necessity of massive-scale data by achieving state-of-the-art performance using only 12,000 whole-slide images from the public TCGA dataset. Built on a ViT-g/14 architecture with 1.1 billion parameters and trained using an optimized DINOv2 recipe, it demonstrates that strategic training techniques can yield high-performance models without requiring millions of proprietary slides. [\href{https://huggingface.co/kaiko-ai/midnight}{https://huggingface.co/kaiko-ai/midnight}]

\textbf{Kaiko \cite{aben2024towards}}: Kaiko refers to the proprietary foundation models developed by Kaiko.ai, which are designed to be robust and generalizable by leveraging large-scale internal pathology datasets. These models typically serve as strong industrial baselines in comparative benchmarks, focusing on clinical-grade performance and robustness across varied scanning conditions and tissue types. [\href{https://huggingface.co/collections/1aurent/kaikoai-models-66636c99d8e1e34bc6dcf795}{https://huggingface.co/collections/1aurent/kaikoai-models-66636c99d8e1e34bc6dcf795}]

\textbf{Lunit \cite{kang2022benchmarking}}: Lunit refers to the set of self-supervised learning models released by Lunit Inc.. Trained on a combination of TCGA and proprietary TULIP datasets, these models provided early and critical insights into the effectiveness of domain-aligned pre-training for computational pathology tasks compared to ImageNet initialization. [\href{https://github.com/lunit-io/benchmark-ssl-pathology}{https://github.com/lunit-io/benchmark-ssl-pathology}]

\textbf{Hibou \cite{nechaev2024hibou}}: Hibou utilizes the DINOv2 framework to train a ViT-L/14 architecture with 304 million parameters on a substantial dataset of 1.1 million whole-slide images. By leveraging over 1.2 billion patches from diverse sources including H\&E and other stains, Hibou-L aims to capture fine-grained histological features, positioning itself as a high-performance open-weights model in the billion-scale data regime. [\href{https://huggingface.co/histai/hibou-L}{https://huggingface.co/histai/hibou-L}]

\begin{table*}[htbp]
\centering
\caption{Overview of selected Pathology Foundation Models.}
\label{tab:pfm_overview}
\resizebox{\textwidth}{!}{%
\begin{tabular}{l|c|c|c|c|c|c|c|c|c|c}
\toprule
\textbf{PFM} & \textbf{Embed Dim} & \textbf{Slides} & \textbf{Patches} & \textbf{Params.} & \textbf{Architecture} & \textbf{Pretraining Strategy} & \textbf{Pretraining Data Source} & \textbf{Stain} & \textbf{Released Date} & \textbf{Publication} \\ \midrule
\multicolumn{11}{c}{\textit{Vision-only}} \\ \midrule
PathOrchestra \cite{yan2025pathorchestra} & 1024 & 300K & - & 303M& ViT-L/16 & DINOv2 & in-house & H\&E & Mar-25 & npj Digital Medicine \\
UNI \cite{chen2024towards} & 1024 & 100,426 & 100M & 303M & ViT-L/16 & DINOv2 & GTEx, In-house & H\&E & 29-Aug-23 & NM \\
UNI2-h \cite{chen2024towards} & 1536 & 350K & 200M & 681M & ViT-H/14 & DINOv2 & GTEx, In-house & H\&E, IHC & 14-Jan-25 & NM \\
Virchow \cite{vorontsov2024foundation} & 1280 & 1,488,550 & 2.0B & 631M & ViT-H/14 & DINOv2 & In-house & H\&E & 14-Sep-23 & NM \\
Virchow2 \cite{zimmermann2024virchow2} & 1280 & 3,134,922 & 2.0B & 631M & ViT-H/14 & DINOv2 & In-house & H\&E, IHC & 1-Aug-24 & Preprint \\
Phikon \cite{Filiot2023ScalingSSLforHistoWithMIM} & 768 & 6,093 & 43.4M & 86.4M & ViT-B/16 & iBOT & TCGA & H\&E & 26-Jul-23 & Preprint \\
\bottomrule
\end{tabular}%
}
\end{table*}

\clearpage 
\section{Frozen PFM Performance }
\renewcommand{\thefigure}{C\arabic{figure}}
\renewcommand{\thetable}{C\arabic{table}}
\setcounter{figure}{0}
\setcounter{table}{0}
% ================= BCSS =================
% \begin{table}[htbp]
% \centering
% \caption{Segmentation performance on \textbf{BCSS} under Frozen PFM. Mean with 95\% confidence interval.}
% \label{tab:bcss}
% \resizebox{\textwidth}{!}{
% \scriptsize
% %
%
}

\vspace{-8pt}
\end{table}
\vspace{-10pt}

% ================= COSAS24 =================
\begin{table}[htbp]
\centering
\caption{Segmentation performance on \textbf{COSAS24} under Frozen PFM. Mean with 95\% confidence interval.}
\label{tab:cosas24}
\resizebox{\textwidth}{!}{
\scriptsize
\begin{tabular}{m{2cm}cccccccc}
\toprule
\textbf{Model} & mIoU & Pixel Acc & mAcc & FW IoU & mDice & mPrecision & mRecall & mF1 \\
\midrule
\shortstack[c]{\textbf{PathOrchestra} \\ \phantom{\tiny [0.000, 0.000]}} & \shortstack[c]{0.670 \\ {\tiny [0.590, 0.748]}} & \shortstack[c]{0.806 \\ {\tiny [0.747, 0.858]}} & \shortstack[c]{0.812 \\ {\tiny [0.763, 0.858]}} & \shortstack[c]{0.681 \\ {\tiny [0.608, 0.753]}} & \shortstack[c]{0.801 \\ {\tiny [0.741, 0.856]}} & \shortstack[c]{0.800 \\ {\tiny [0.741, 0.854]}} & \shortstack[c]{0.812 \\ {\tiny [0.763, 0.858]}} & \shortstack[c]{0.801 \\ {\tiny [0.741, 0.856]}} \\
\shortstack[c]{\textbf{CONCHv1.5} \\ \phantom{\tiny [0.000, 0.000]}} & \shortstack[c]{0.719 \\ {\tiny [0.654, 0.786]}} & \shortstack[c]{0.840 \\ {\tiny [0.796, 0.882]}} & \shortstack[c]{0.850 \\ {\tiny [0.809, 0.888]}} & \shortstack[c]{0.728 \\ {\tiny [0.667, 0.790]}} & \shortstack[c]{0.836 \\ {\tiny [0.789, 0.880]}} & \shortstack[c]{0.835 \\ {\tiny [0.787, 0.880]}} & \shortstack[c]{0.850 \\ {\tiny [0.809, 0.888]}} & \shortstack[c]{0.836 \\ {\tiny [0.789, 0.880]}} \\
\shortstack[c]{\textbf{CONCH} \\ \phantom{\tiny [0.000, 0.000]}} & \shortstack[c]{0.744 \\ {\tiny [0.694, 0.796]}} & \shortstack[c]{0.857 \\ {\tiny [0.826, 0.887]}} & \shortstack[c]{0.865 \\ {\tiny [0.838, 0.892]}} & \shortstack[c]{0.753 \\ {\tiny [0.709, 0.798]}} & \shortstack[c]{0.853 \\ {\tiny [0.817, 0.886]}} & \shortstack[c]{0.850 \\ {\tiny [0.811, 0.885]}} & \shortstack[c]{0.865 \\ {\tiny [0.838, 0.892]}} & \shortstack[c]{0.853 \\ {\tiny [0.817, 0.886]}} \\
\bottomrule
\end{tabular}}

\vspace{-8pt}
\end{table}
\vspace{-10pt}

% ================= CRAG =================
\begin{table}[htbp]
\centering
\caption{Segmentation performance on \textbf{CRAG} under Frozen PFM. Mean with 95\% confidence interval.}
\label{tab:crag}
\resizebox{\textwidth}{!}{
\scriptsize
\begin{tabular}{m{2cm}cccccccc}
\toprule
\textbf{Model} & mIoU & Pixel Acc & mAcc & FW IoU & mDice & mPrecision & mRecall & mF1 \\
\midrule
\shortstack[c]{\textbf{PathOrchestra} \\ \phantom{\tiny [0.000, 0.000]}} & \shortstack[c]{0.827 \\ {\tiny [0.771, 0.874]}} & \shortstack[c]{0.910 \\ {\tiny [0.875, 0.940]}} & \shortstack[c]{0.920 \\ {\tiny [0.889, 0.943]}} & \shortstack[c]{0.839 \\ {\tiny [0.780, 0.889]}} & \shortstack[c]{0.905 \\ {\tiny [0.870, 0.932]}} & \shortstack[c]{0.897 \\ {\tiny [0.864, 0.927]}} & \shortstack[c]{0.920 \\ {\tiny [0.889, 0.943]}} & \shortstack[c]{0.905 \\ {\tiny [0.870, 0.932]}} \\
\shortstack[c]{\textbf{CONCHv1.5} \\ \phantom{\tiny [0.000, 0.000]}} & \shortstack[c]{0.859 \\ {\tiny [0.825, 0.890]}} & \shortstack[c]{0.930 \\ {\tiny [0.908, 0.950]}} & \shortstack[c]{0.925 \\ {\tiny [0.908, 0.941]}} & \shortstack[c]{0.870 \\ {\tiny [0.833, 0.905]}} & \shortstack[c]{0.924 \\ {\tiny [0.904, 0.941]}} & \shortstack[c]{0.922 \\ {\tiny [0.901, 0.943]}} & \shortstack[c]{0.925 \\ {\tiny [0.908, 0.941]}} & \shortstack[c]{0.924 \\ {\tiny [0.904, 0.941]}} \\
\shortstack[c]{\textbf{CONCH} \\ \phantom{\tiny [0.000, 0.000]}} & \shortstack[c]{0.846 \\ {\tiny [0.807, 0.882]}} & \shortstack[c]{0.922 \\ {\tiny [0.897, 0.945]}} & \shortstack[c]{0.922 \\ {\tiny [0.900, 0.942]}} & \shortstack[c]{0.858 \\ {\tiny [0.815, 0.898]}} & \shortstack[c]{0.916 \\ {\tiny [0.893, 0.936]}} & \shortstack[c]{0.911 \\ {\tiny [0.887, 0.933]}} & \shortstack[c]{0.922 \\ {\tiny [0.900, 0.942]}} & \shortstack[c]{0.916 \\ {\tiny [0.893, 0.936]}} \\
\bottomrule
\end{tabular}
}
\vspace{-8pt}
\end{table}
\vspace{-10pt}

% ================= CoCaHis =================
\begin{table}[htbp]
\centering
\caption{Segmentation performance on \textbf{CoCaHis} under Frozen PFM. Mean with 95\% confidence interval.}
\label{tab:cocahis}
\resizebox{\textwidth}{!}{
\scriptsize
\begin{tabular}{m{2cm}cccccccc}
\toprule
\textbf{Model} & mIoU & Pixel Acc & mAcc & FW IoU & mDice & mPrecision & mRecall & mF1 \\
\midrule
\shortstack[c]{\textbf{PathOrchestra} \\ \phantom{\tiny [0.000, 0.000]}} & \shortstack[c]{0.813 \\ {\tiny [0.775, 0.850]}} & \shortstack[c]{0.904 \\ {\tiny [0.875, 0.931]}} & \shortstack[c]{0.895 \\ {\tiny [0.870, 0.922]}} & \shortstack[c]{0.826 \\ {\tiny [0.777, 0.873]}} & \shortstack[c]{0.896 \\ {\tiny [0.873, 0.919]}} & \shortstack[c]{0.898 \\ {\tiny [0.876, 0.916]}} & \shortstack[c]{0.895 \\ {\tiny [0.870, 0.922]}} & \shortstack[c]{0.896 \\ {\tiny [0.873, 0.919]}} \\
\shortstack[c]{\textbf{CONCHv1.5} \\ \phantom{\tiny [0.000, 0.000]}} & \shortstack[c]{0.841 \\ {\tiny [0.810, 0.868]}} & \shortstack[c]{0.919 \\ {\tiny [0.896, 0.940]}} & \shortstack[c]{0.919 \\ {\tiny [0.897, 0.935]}} & \shortstack[c]{0.852 \\ {\tiny [0.810, 0.889]}} & \shortstack[c]{0.913 \\ {\tiny [0.895, 0.929]}} & \shortstack[c]{0.910 \\ {\tiny [0.895, 0.925]}} & \shortstack[c]{0.919 \\ {\tiny [0.897, 0.935]}} & \shortstack[c]{0.913 \\ {\tiny [0.895, 0.929]}} \\
\shortstack[c]{\textbf{CONCH} \\ \phantom{\tiny [0.000, 0.000]}} & \shortstack[c]{0.835 \\ {\tiny [0.807, 0.861]}} & \shortstack[c]{0.917 \\ {\tiny [0.894, 0.937]}} & \shortstack[c]{0.909 \\ {\tiny [0.893, 0.924]}} & \shortstack[c]{0.847 \\ {\tiny [0.808, 0.883]}} & \shortstack[c]{0.910 \\ {\tiny [0.893, 0.925]}} & \shortstack[c]{0.911 \\ {\tiny [0.893, 0.927]}} & \shortstack[c]{0.909 \\ {\tiny [0.893, 0.924]}} & \shortstack[c]{0.910 \\ {\tiny [0.893, 0.925]}} \\
\bottomrule
\end{tabular}
}
\vspace{-8pt}
\end{table}
\vspace{-10pt}

% ================= CoNIC2022 =================
\begin{table}[htbp]
\centering
\caption{Segmentation performance on \textbf{CoNIC2022} under Frozen PFM. Mean with 95\% confidence interval.}
\label{tab:conic2022}
\resizebox{\textwidth}{!}{
\scriptsize
\begin{tabular}{m{2cm}cccccccc}
\toprule
\textbf{Model} & mIoU & Pixel Acc & mAcc & FW IoU & mDice & mPrecision & mRecall & mF1 \\
\midrule
\shortstack[c]{\textbf{PathOrchestra} \\ \phantom{\tiny [0.000, 0.000]}} & \shortstack[c]{0.358 \\ {\tiny [0.351, 0.365]}} & \shortstack[c]{0.875 \\ {\tiny [0.869, 0.880]}} & \shortstack[c]{0.555 \\ {\tiny [0.542, 0.568]}} & \shortstack[c]{0.803 \\ {\tiny [0.795, 0.812]}} & \shortstack[c]{0.488 \\ {\tiny [0.478, 0.498]}} & \shortstack[c]{0.449 \\ {\tiny [0.440, 0.458]}} & \shortstack[c]{0.555 \\ {\tiny [0.542, 0.568]}} & \shortstack[c]{0.488 \\ {\tiny [0.478, 0.498]}} \\
\shortstack[c]{\textbf{CONCHv1.5} \\ \phantom{\tiny [0.000, 0.000]}} & \shortstack[c]{0.361 \\ {\tiny [0.354, 0.368]}} & \shortstack[c]{0.878 \\ {\tiny [0.872, 0.883]}} & \shortstack[c]{0.546 \\ {\tiny [0.533, 0.559]}} & \shortstack[c]{0.807 \\ {\tiny [0.798, 0.815]}} & \shortstack[c]{0.492 \\ {\tiny [0.481, 0.502]}} & \shortstack[c]{0.458 \\ {\tiny [0.448, 0.468]}} & \shortstack[c]{0.546 \\ {\tiny [0.533, 0.559]}} & \shortstack[c]{0.492 \\ {\tiny [0.481, 0.502]}} \\
\shortstack[c]{\textbf{CONCH} \\ \phantom{\tiny [0.000, 0.000]}} & \shortstack[c]{0.356 \\ {\tiny [0.349, 0.363]}} & \shortstack[c]{0.871 \\ {\tiny [0.865, 0.877]}} & \shortstack[c]{0.563 \\ {\tiny [0.549, 0.576]}} & \shortstack[c]{0.799 \\ {\tiny [0.790, 0.808]}} & \shortstack[c]{0.486 \\ {\tiny [0.476, 0.496]}} & \shortstack[c]{0.441 \\ {\tiny [0.433, 0.450]}} & \shortstack[c]{0.563 \\ {\tiny [0.549, 0.576]}} & \shortstack[c]{0.486 \\ {\tiny [0.476, 0.496]}} \\
\bottomrule
\end{tabular}
}
\vspace{-8pt}
\end{table}
\vspace{-10pt}

% ================= CoNSeP =================
\begin{table}[htbp]
\centering
\caption{Segmentation performance on \textbf{CoNSeP} under Frozen PFM. Mean with 95\% confidence interval.}
\label{tab:consep}
\resizebox{\textwidth}{!}{
\scriptsize
\begin{tabular}{m{2cm}cccccccc}
\toprule
\textbf{Model} & mIoU & Pixel Acc & mAcc & FW IoU & mDice & mPrecision & mRecall & mF1 \\
\midrule
\shortstack[c]{\textbf{PathOrchestra} \\ \phantom{\tiny [0.000, 0.000]}} & \shortstack[c]{0.208 \\ {\tiny [0.152, 0.306]}} & \shortstack[c]{0.867 \\ {\tiny [0.809, 0.921]}} & \shortstack[c]{0.314 \\ {\tiny [0.218, 0.444]}} & \shortstack[c]{0.808 \\ {\tiny [0.744, 0.876]}} & \shortstack[c]{0.267 \\ {\tiny [0.185, 0.392]}} & \shortstack[c]{0.464 \\ {\tiny [0.380, 0.564]}} & \shortstack[c]{0.314 \\ {\tiny [0.218, 0.444]}} & \shortstack[c]{0.521 \\ {\tiny [0.450, 0.590]}} \\
\shortstack[c]{\textbf{CONCHv1.5} \\ \phantom{\tiny [0.000, 0.000]}} & \shortstack[c]{0.341 \\ {\tiny [0.225, 0.435]}} & \shortstack[c]{0.904 \\ {\tiny [0.871, 0.940]}} & \shortstack[c]{0.523 \\ {\tiny [0.382, 0.675]}} & \shortstack[c]{0.853 \\ {\tiny [0.803, 0.910]}} & \shortstack[c]{0.439 \\ {\tiny [0.287, 0.557]}} & \shortstack[c]{0.531 \\ {\tiny [0.408, 0.613]}} & \shortstack[c]{0.523 \\ {\tiny [0.382, 0.675]}} & \shortstack[c]{0.617 \\ {\tiny [0.568, 0.656]}} \\
\shortstack[c]{\textbf{CONCH} \\ \phantom{\tiny [0.000, 0.000]}} & \shortstack[c]{0.279 \\ {\tiny [0.219, 0.359]}} & \shortstack[c]{0.885 \\ {\tiny [0.834, 0.936]}} & \shortstack[c]{0.451 \\ {\tiny [0.353, 0.585]}} & \shortstack[c]{0.838 \\ {\tiny [0.775, 0.905]}} & \shortstack[c]{0.366 \\ {\tiny [0.281, 0.458]}} & \shortstack[c]{0.398 \\ {\tiny [0.339, 0.436]}} & \shortstack[c]{0.451 \\ {\tiny [0.353, 0.585]}} & \shortstack[c]{0.525 \\ {\tiny [0.453, 0.641]}} \\
\bottomrule
\end{tabular}
}
\vspace{-8pt}
\end{table}
\vspace{-10pt}

% ================= EBHI =================
\begin{table}[htbp]
\centering
\caption{Segmentation performance on \textbf{EBHI} under Frozen PFM. Mean with 95\% confidence interval.}
\label{tab:ebhi}
\resizebox{\textwidth}{!}{
\scriptsize
\begin{tabular}{m{2cm}cccccccc}
\toprule
\textbf{Model} & mIoU & Pixel Acc & mAcc & FW IoU & mDice & mPrecision & mRecall & mF1 \\
\midrule
\shortstack[c]{\textbf{PathOrchestra} \\ \phantom{\tiny [0.000, 0.000]}} & \shortstack[c]{0.426 \\ {\tiny [0.404, 0.445]}} & \shortstack[c]{0.820 \\ {\tiny [0.786, 0.850]}} & \shortstack[c]{0.524 \\ {\tiny [0.519, 0.528]}} & \shortstack[c]{0.689 \\ {\tiny [0.639, 0.734]}} & \shortstack[c]{0.487 \\ {\tiny [0.472, 0.500]}} & \shortstack[c]{0.803 \\ {\tiny [0.763, 0.839]}} & \shortstack[c]{0.524 \\ {\tiny [0.519, 0.528]}} & \shortstack[c]{0.852 \\ {\tiny [0.826, 0.875]}} \\
\shortstack[c]{\textbf{CONCHv1.5} \\ \phantom{\tiny [0.000, 0.000]}} & \shortstack[c]{0.414 \\ {\tiny [0.393, 0.435]}} & \shortstack[c]{0.809 \\ {\tiny [0.776, 0.839]}} & \shortstack[c]{0.514 \\ {\tiny [0.505, 0.521]}} & \shortstack[c]{0.672 \\ {\tiny [0.625, 0.718]}} & \shortstack[c]{0.479 \\ {\tiny [0.464, 0.493]}} & \shortstack[c]{0.789 \\ {\tiny [0.749, 0.825]}} & \shortstack[c]{0.514 \\ {\tiny [0.505, 0.521]}} & \shortstack[c]{0.838 \\ {\tiny [0.811, 0.863]}} \\
\shortstack[c]{\textbf{CONCH} \\ \phantom{\tiny [0.000, 0.000]}} & \shortstack[c]{0.420 \\ {\tiny [0.398, 0.440]}} & \shortstack[c]{0.816 \\ {\tiny [0.782, 0.845]}} & \shortstack[c]{0.517 \\ {\tiny [0.507, 0.526]}} & \shortstack[c]{0.683 \\ {\tiny [0.635, 0.727]}} & \shortstack[c]{0.483 \\ {\tiny [0.467, 0.496]}} & \shortstack[c]{0.798 \\ {\tiny [0.757, 0.834]}} & \shortstack[c]{0.517 \\ {\tiny [0.507, 0.526]}} & \shortstack[c]{0.845 \\ {\tiny [0.817, 0.868]}} \\
\bottomrule
\end{tabular}
}
\vspace{-8pt}
\end{table}
\vspace{-10pt}

% ================= GlaS =================
\begin{table}[htbp]
\centering
\caption{Segmentation performance on \textbf{GlaS} under Frozen PFM. Mean with 95\% confidence interval.}
\label{tab:glas}
\resizebox{\textwidth}{!}{
\scriptsize
\begin{tabular}{m{2cm}cccccccc}
\toprule
\textbf{Model} & mIoU & Pixel Acc & mAcc & FW IoU & mDice & mPrecision & mRecall & mF1 \\
\midrule
\shortstack[c]{\textbf{PathOrchestra} \\ \phantom{\tiny [0.000, 0.000]}} & \shortstack[c]{0.880 \\ {\tiny [0.857, 0.900]}} & \shortstack[c]{0.937 \\ {\tiny [0.924, 0.948]}} & \shortstack[c]{0.937 \\ {\tiny [0.923, 0.949]}} & \shortstack[c]{0.881 \\ {\tiny [0.858, 0.902]}} & \shortstack[c]{0.936 \\ {\tiny [0.923, 0.947]}} & \shortstack[c]{0.936 \\ {\tiny [0.923, 0.947]}} & \shortstack[c]{0.937 \\ {\tiny [0.923, 0.949]}} & \shortstack[c]{0.936 \\ {\tiny [0.923, 0.947]}} \\
\shortstack[c]{\textbf{CONCHv1.5} \\ \phantom{\tiny [0.000, 0.000]}} & \shortstack[c]{0.877 \\ {\tiny [0.855, 0.898]}} & \shortstack[c]{0.935 \\ {\tiny [0.922, 0.947]}} & \shortstack[c]{0.936 \\ {\tiny [0.921, 0.949]}} & \shortstack[c]{0.878 \\ {\tiny [0.856, 0.900]}} & \shortstack[c]{0.935 \\ {\tiny [0.922, 0.946]}} & \shortstack[c]{0.934 \\ {\tiny [0.923, 0.945]}} & \shortstack[c]{0.936 \\ {\tiny [0.921, 0.949]}} & \shortstack[c]{0.935 \\ {\tiny [0.922, 0.946]}} \\
\shortstack[c]{\textbf{CONCH} \\ \phantom{\tiny [0.000, 0.000]}} & \shortstack[c]{0.875 \\ {\tiny [0.852, 0.897]}} & \shortstack[c]{0.934 \\ {\tiny [0.920, 0.947]}} & \shortstack[c]{0.935 \\ {\tiny [0.919, 0.948]}} & \shortstack[c]{0.876 \\ {\tiny [0.852, 0.900]}} & \shortstack[c]{0.934 \\ {\tiny [0.920, 0.946]}} & \shortstack[c]{0.933 \\ {\tiny [0.922, 0.944]}} & \shortstack[c]{0.935 \\ {\tiny [0.919, 0.948]}} & \shortstack[c]{0.934 \\ {\tiny [0.920, 0.946]}} \\
\bottomrule
\end{tabular}
}
\vspace{-8pt}
\end{table}
\vspace{-10pt}

% ================= Janowczyk =================
\begin{table}[htbp]
\centering
\caption{Segmentation performance on \textbf{Janowczyk} under Frozen PFM. Mean with 95\% confidence interval.}
\label{tab:janowczyk}
\resizebox{\textwidth}{!}{
\scriptsize
\begin{tabular}{m{2cm}cccccccc}
\toprule
\textbf{Model} & mIoU & Pixel Acc & mAcc & FW IoU & mDice & mPrecision & mRecall & mF1 \\
\midrule
\shortstack[c]{\textbf{PathOrchestra} \\ \phantom{\tiny [0.000, 0.000]}} & \shortstack[c]{0.560 \\ {\tiny [0.541, 0.577]}} & \shortstack[c]{0.975 \\ {\tiny [0.970, 0.979]}} & \shortstack[c]{0.602 \\ {\tiny [0.570, 0.631]}} & \shortstack[c]{0.958 \\ {\tiny [0.948, 0.966]}} & \shortstack[c]{0.620 \\ {\tiny [0.589, 0.646]}} & \shortstack[c]{0.651 \\ {\tiny [0.614, 0.692]}} & \shortstack[c]{0.602 \\ {\tiny [0.570, 0.631]}} & \shortstack[c]{0.620 \\ {\tiny [0.589, 0.646]}} \\
\shortstack[c]{\textbf{CONCHv1.5} \\ \phantom{\tiny [0.000, 0.000]}} & \shortstack[c]{0.579 \\ {\tiny [0.561, 0.595]}} & \shortstack[c]{0.972 \\ {\tiny [0.966, 0.978]}} & \shortstack[c]{0.650 \\ {\tiny [0.613, 0.681]}} & \shortstack[c]{0.956 \\ {\tiny [0.946, 0.965]}} & \shortstack[c]{0.649 \\ {\tiny [0.621, 0.673]}} & \shortstack[c]{0.650 \\ {\tiny [0.624, 0.683]}} & \shortstack[c]{0.650 \\ {\tiny [0.613, 0.681]}} & \shortstack[c]{0.649 \\ {\tiny [0.621, 0.673]}} \\
\shortstack[c]{\textbf{CONCH} \\ \phantom{\tiny [0.000, 0.000]}} & \shortstack[c]{0.574 \\ {\tiny [0.552, 0.594]}} & \shortstack[c]{0.966 \\ {\tiny [0.961, 0.971]}} & \shortstack[c]{0.678 \\ {\tiny [0.619, 0.733]}} & \shortstack[c]{0.950 \\ {\tiny [0.942, 0.957]}} & \shortstack[c]{0.645 \\ {\tiny [0.612, 0.673]}} & \shortstack[c]{0.624 \\ {\tiny [0.596, 0.657]}} & \shortstack[c]{0.678 \\ {\tiny [0.619, 0.733]}} & \shortstack[c]{0.645 \\ {\tiny [0.612, 0.673]}} \\
\bottomrule
\end{tabular}
}
\vspace{-8pt}
\end{table}
\vspace{-10pt}

% ================= Lizard =================
\begin{table}[htbp]
\centering
\caption{Segmentation performance on \textbf{Lizard} under Frozen PFM. Mean with 95\% confidence interval.}
\label{tab:lizard}
\resizebox{\textwidth}{!}{
\scriptsize
\begin{tabular}{m{2cm}cccccccc}
\toprule
\textbf{Model} & mIoU & Pixel Acc & mAcc & FW IoU & mDice & mPrecision & mRecall & mF1 \\
\midrule
\shortstack[c]{\textbf{PathOrchestra} \\ \phantom{\tiny [0.000, 0.000]}} & \shortstack[c]{0.207 \\ {\tiny [0.185, 0.227]}} & \shortstack[c]{0.788 \\ {\tiny [0.762, 0.815]}} & \shortstack[c]{0.307 \\ {\tiny [0.266, 0.341]}} & \shortstack[c]{0.695 \\ {\tiny [0.658, 0.733]}} & \shortstack[c]{0.280 \\ {\tiny [0.246, 0.307]}} & \shortstack[c]{0.370 \\ {\tiny [0.328, 0.405]}} & \shortstack[c]{0.307 \\ {\tiny [0.266, 0.341]}} & \shortstack[c]{0.392 \\ {\tiny [0.345, 0.430]}} \\
\shortstack[c]{\textbf{CONCHv1.5} \\ \phantom{\tiny [0.000, 0.000]}} & \shortstack[c]{0.217 \\ {\tiny [0.195, 0.238]}} & \shortstack[c]{0.789 \\ {\tiny [0.760, 0.820]}} & \shortstack[c]{0.326 \\ {\tiny [0.285, 0.359]}} & \shortstack[c]{0.697 \\ {\tiny [0.659, 0.738]}} & \shortstack[c]{0.294 \\ {\tiny [0.261, 0.322]}} & \shortstack[c]{0.385 \\ {\tiny [0.344, 0.421]}} & \shortstack[c]{0.326 \\ {\tiny [0.285, 0.359]}} & \shortstack[c]{0.412 \\ {\tiny [0.366, 0.451]}} \\
\shortstack[c]{\textbf{CONCH} \\ \phantom{\tiny [0.000, 0.000]}} & \shortstack[c]{0.214 \\ {\tiny [0.191, 0.235]}} & \shortstack[c]{0.788 \\ {\tiny [0.758, 0.818]}} & \shortstack[c]{0.323 \\ {\tiny [0.280, 0.360]}} & \shortstack[c]{0.695 \\ {\tiny [0.655, 0.736]}} & \shortstack[c]{0.289 \\ {\tiny [0.256, 0.319]}} & \shortstack[c]{0.336 \\ {\tiny [0.284, 0.406]}} & \shortstack[c]{0.323 \\ {\tiny [0.280, 0.360]}} & \shortstack[c]{0.405 \\ {\tiny [0.358, 0.447]}} \\
\bottomrule
\end{tabular}
}
\vspace{-8pt}
\end{table}
\vspace{-10pt}

% ================= NuCLS =================
\begin{table}[htbp]
\centering
\caption{Segmentation performance on \textbf{NuCLS} under Frozen PFM. Mean with 95\% confidence interval.}
\label{tab:nucls}
\resizebox{\textwidth}{!}{
\scriptsize
\begin{tabular}{m{2cm}cccccccc}
\toprule
\textbf{Model} & mIoU & Pixel Acc & mAcc & FW IoU & mDice & mPrecision & mRecall & mF1 \\
\midrule
\shortstack[c]{\textbf{PathOrchestra} \\ \phantom{\tiny [0.000, 0.000]}} & \shortstack[c]{0.213 \\ {\tiny [0.195, 0.253]}} & \shortstack[c]{0.788 \\ {\tiny [0.777, 0.799]}} & \shortstack[c]{0.320 \\ {\tiny [0.293, 0.380]}} & \shortstack[c]{0.683 \\ {\tiny [0.669, 0.697]}} & \shortstack[c]{0.287 \\ {\tiny [0.264, 0.340]}} & \shortstack[c]{0.477 \\ {\tiny [0.456, 0.498]}} & \shortstack[c]{0.320 \\ {\tiny [0.293, 0.380]}} & \shortstack[c]{0.502 \\ {\tiny [0.481, 0.523]}} \\
\shortstack[c]{\textbf{CONCHv1.5} \\ \phantom{\tiny [0.000, 0.000]}} & \shortstack[c]{0.221 \\ {\tiny [0.203, 0.262]}} & \shortstack[c]{0.796 \\ {\tiny [0.785, 0.807]}} & \shortstack[c]{0.321 \\ {\tiny [0.294, 0.380]}} & \shortstack[c]{0.692 \\ {\tiny [0.677, 0.706]}} & \shortstack[c]{0.298 \\ {\tiny [0.274, 0.353]}} & \shortstack[c]{0.495 \\ {\tiny [0.473, 0.517]}} & \shortstack[c]{0.321 \\ {\tiny [0.294, 0.380]}} & \shortstack[c]{0.520 \\ {\tiny [0.498, 0.542]}} \\
\shortstack[c]{\textbf{CONCH} \\ \phantom{\tiny [0.000, 0.000]}} & \shortstack[c]{0.219 \\ {\tiny [0.201, 0.258]}} & \shortstack[c]{0.798 \\ {\tiny [0.788, 0.808]}} & \shortstack[c]{0.315 \\ {\tiny [0.291, 0.373]}} & \shortstack[c]{0.693 \\ {\tiny [0.680, 0.707]}} & \shortstack[c]{0.294 \\ {\tiny [0.271, 0.347]}} & \shortstack[c]{0.491 \\ {\tiny [0.470, 0.514]}} & \shortstack[c]{0.315 \\ {\tiny [0.291, 0.373]}} & \shortstack[c]{0.514 \\ {\tiny [0.494, 0.533]}} \\
\bottomrule
\end{tabular}
}
\vspace{-8pt}
\end{table}
\vspace{-10pt}

% ================= PanNuke =================
\begin{table}[htbp]
\centering
\caption{Segmentation performance on \textbf{PanNuke} under Frozen PFM. Mean with 95\% confidence interval.}
\label{tab:pannuke}
\resizebox{\textwidth}{!}{
\scriptsize
\begin{tabular}{m{2cm}cccccccc}
\toprule
\textbf{Model} & mIoU & Pixel Acc & mAcc & FW IoU & mDice & mPrecision & mRecall & mF1 \\
\midrule
\shortstack[c]{\textbf{PathOrchestra} \\ \phantom{\tiny [0.000, 0.000]}} & \shortstack[c]{0.461 \\ {\tiny [0.451, 0.471]}} & \shortstack[c]{0.897 \\ {\tiny [0.889, 0.904]}} & \shortstack[c]{0.584 \\ {\tiny [0.575, 0.594]}} & \shortstack[c]{0.829 \\ {\tiny [0.816, 0.840]}} & \shortstack[c]{0.580 \\ {\tiny [0.571, 0.588]}} & \shortstack[c]{0.692 \\ {\tiny [0.680, 0.704]}} & \shortstack[c]{0.584 \\ {\tiny [0.575, 0.594]}} & \shortstack[c]{0.696 \\ {\tiny [0.685, 0.706]}} \\
\shortstack[c]{\textbf{CONCHv1.5} \\ \phantom{\tiny [0.000, 0.000]}} & \shortstack[c]{0.486 \\ {\tiny [0.476, 0.496]}} & \shortstack[c]{0.907 \\ {\tiny [0.901, 0.913]}} & \shortstack[c]{0.615 \\ {\tiny [0.604, 0.624]}} & \shortstack[c]{0.845 \\ {\tiny [0.835, 0.854]}} & \shortstack[c]{0.601 \\ {\tiny [0.592, 0.609]}} & \shortstack[c]{0.707 \\ {\tiny [0.695, 0.719]}} & \shortstack[c]{0.615 \\ {\tiny [0.604, 0.624]}} & \shortstack[c]{0.721 \\ {\tiny [0.710, 0.731]}} \\
\shortstack[c]{\textbf{CONCH} \\ \phantom{\tiny [0.000, 0.000]}} & \shortstack[c]{0.480 \\ {\tiny [0.469, 0.490]}} & \shortstack[c]{0.906 \\ {\tiny [0.900, 0.912]}} & \shortstack[c]{0.603 \\ {\tiny [0.593, 0.613]}} & \shortstack[c]{0.843 \\ {\tiny [0.832, 0.853]}} & \shortstack[c]{0.595 \\ {\tiny [0.585, 0.603]}} & \shortstack[c]{0.705 \\ {\tiny [0.694, 0.717]}} & \shortstack[c]{0.603 \\ {\tiny [0.593, 0.613]}} & \shortstack[c]{0.714 \\ {\tiny [0.702, 0.724]}} \\
\bottomrule
\end{tabular}
}
\vspace{-8pt}
\end{table}
\vspace{-10pt}

% ================= RINGS =================
\begin{table}[htbp]
\centering
\caption{Segmentation performance on \textbf{RINGS} under Frozen PFM. Mean with 95\% confidence interval.}
\label{tab:rings}
\resizebox{\textwidth}{!}{
\scriptsize
\begin{tabular}{m{2cm}cccccccc}
\toprule
\textbf{Model} & mIoU & Pixel Acc & mAcc & FW IoU & mDice & mPrecision & mRecall & mF1 \\
\midrule
\shortstack[c]{\textbf{PathOrchestra} \\ \phantom{\tiny [0.000, 0.000]}} & \shortstack[c]{0.883 \\ {\tiny [0.869, 0.897]}} & \shortstack[c]{0.947 \\ {\tiny [0.941, 0.954]}} & \shortstack[c]{0.940 \\ {\tiny [0.931, 0.948]}} & \shortstack[c]{0.901 \\ {\tiny [0.889, 0.913]}} & \shortstack[c]{0.938 \\ {\tiny [0.929, 0.945]}} & \shortstack[c]{0.935 \\ {\tiny [0.927, 0.944]}} & \shortstack[c]{0.940 \\ {\tiny [0.931, 0.948]}} & \shortstack[c]{0.938 \\ {\tiny [0.929, 0.945]}} \\
\shortstack[c]{\textbf{CONCHv1.5} \\ \phantom{\tiny [0.000, 0.000]}} & \shortstack[c]{0.887 \\ {\tiny [0.874, 0.901]}} & \shortstack[c]{0.950 \\ {\tiny [0.943, 0.956]}} & \shortstack[c]{0.938 \\ {\tiny [0.930, 0.947]}} & \shortstack[c]{0.905 \\ {\tiny [0.892, 0.917]}} & \shortstack[c]{0.940 \\ {\tiny [0.932, 0.948]}} & \shortstack[c]{0.941 \\ {\tiny [0.933, 0.949]}} & \shortstack[c]{0.938 \\ {\tiny [0.930, 0.947]}} & \shortstack[c]{0.940 \\ {\tiny [0.932, 0.948]}} \\
\shortstack[c]{\textbf{CONCH} \\ \phantom{\tiny [0.000, 0.000]}} & \shortstack[c]{0.884 \\ {\tiny [0.869, 0.898]}} & \shortstack[c]{0.948 \\ {\tiny [0.942, 0.955]}} & \shortstack[c]{0.939 \\ {\tiny [0.930, 0.947]}} & \shortstack[c]{0.902 \\ {\tiny [0.891, 0.915]}} & \shortstack[c]{0.938 \\ {\tiny [0.929, 0.946]}} & \shortstack[c]{0.938 \\ {\tiny [0.927, 0.946]}} & \shortstack[c]{0.939 \\ {\tiny [0.930, 0.947]}} & \shortstack[c]{0.938 \\ {\tiny [0.929, 0.946]}} \\
\bottomrule
\end{tabular}
}
\vspace{-8pt}
\end{table}
\vspace{-10pt}

% ================= TNBC =================
\begin{table}[htbp]
\centering
\caption{Segmentation performance on \textbf{TNBC} under Frozen PFM. Mean with 95\% confidence interval.}
\label{tab:tnbc}
\resizebox{\textwidth}{!}{
\scriptsize
\begin{tabular}{m{2cm}cccccccc}
\toprule
\textbf{Model} & mIoU & Pixel Acc & mAcc & FW IoU & mDice & mPrecision & mRecall & mF1 \\
\midrule
\shortstack[c]{\textbf{PathOrchestra} \\ \phantom{\tiny [0.000, 0.000]}} & \shortstack[c]{0.720 \\ {\tiny [0.678, 0.766]}} & \shortstack[c]{0.939 \\ {\tiny [0.912, 0.967]}} & \shortstack[c]{0.830 \\ {\tiny [0.769, 0.868]}} & \shortstack[c]{0.896 \\ {\tiny [0.857, 0.944]}} & \shortstack[c]{0.819 \\ {\tiny [0.781, 0.855]}} & \shortstack[c]{0.810 \\ {\tiny [0.764, 0.850]}} & \shortstack[c]{0.830 \\ {\tiny [0.769, 0.868]}} & \shortstack[c]{0.819 \\ {\tiny [0.781, 0.855]}} \\
\shortstack[c]{\textbf{CONCHv1.5} \\ \phantom{\tiny [0.000, 0.000]}} & \shortstack[c]{0.729 \\ {\tiny [0.686, 0.771]}} & \shortstack[c]{0.938 \\ {\tiny [0.909, 0.968]}} & \shortstack[c]{0.852 \\ {\tiny [0.820, 0.883]}} & \shortstack[c]{0.897 \\ {\tiny [0.854, 0.945]}} & \shortstack[c]{0.826 \\ {\tiny [0.793, 0.859]}} & \shortstack[c]{0.806 \\ {\tiny [0.763, 0.845]}} & \shortstack[c]{0.852 \\ {\tiny [0.820, 0.883]}} & \shortstack[c]{0.826 \\ {\tiny [0.793, 0.859]}} \\
\shortstack[c]{\textbf{CONCH} \\ \phantom{\tiny [0.000, 0.000]}} & \shortstack[c]{0.716 \\ {\tiny [0.682, 0.755]}} & \shortstack[c]{0.934 \\ {\tiny [0.910, 0.964]}} & \shortstack[c]{0.847 \\ {\tiny [0.792, 0.899]}} & \shortstack[c]{0.891 \\ {\tiny [0.854, 0.939]}} & \shortstack[c]{0.816 \\ {\tiny [0.785, 0.847]}} & \shortstack[c]{0.793 \\ {\tiny [0.771, 0.822]}} & \shortstack[c]{0.847 \\ {\tiny [0.792, 0.899]}} & \shortstack[c]{0.816 \\ {\tiny [0.785, 0.847]}} \\
\bottomrule
\end{tabular}
}
\vspace{-8pt}
\end{table}
\vspace{-10pt}

% ================= WSSS4LUAD =================
\begin{table}[htbp]
\centering
\caption{Segmentation performance on \textbf{WSSS4LUAD} under Frozen PFM. Mean with 95\% confidence interval.}
\label{tab:wsss4luad}
\resizebox{\textwidth}{!}{
\scriptsize
\begin{tabular}{m{2cm}cccccccc}
\toprule
\textbf{Model} & mIoU & Pixel Acc & mAcc & FW IoU & mDice & mPrecision & mRecall & mF1 \\
\midrule
\shortstack[c]{\textbf{PathOrchestra} \\ \phantom{\tiny [0.000, 0.000]}} & \shortstack[c]{0.413 \\ {\tiny [0.355, 0.469]}} & \shortstack[c]{0.690 \\ {\tiny [0.615, 0.766]}} & \shortstack[c]{0.569 \\ {\tiny [0.502, 0.751]}} & \shortstack[c]{0.528 \\ {\tiny [0.437, 0.629]}} & \shortstack[c]{0.545 \\ {\tiny [0.488, 0.602]}} & \shortstack[c]{0.595 \\ {\tiny [0.502, 0.669]}} & \shortstack[c]{0.569 \\ {\tiny [0.502, 0.751]}} & \shortstack[c]{0.563 \\ {\tiny [0.491, 0.731]}} \\
\shortstack[c]{\textbf{CONCHv1.5} \\ \phantom{\tiny [0.000, 0.000]}} & \shortstack[c]{0.466 \\ {\tiny [0.390, 0.530]}} & \shortstack[c]{0.698 \\ {\tiny [0.614, 0.780]}} & \shortstack[c]{0.649 \\ {\tiny [0.571, 0.747]}} & \shortstack[c]{0.543 \\ {\tiny [0.445, 0.647]}} & \shortstack[c]{0.621 \\ {\tiny [0.515, 0.679]}} & \shortstack[c]{0.627 \\ {\tiny [0.520, 0.708]}} & \shortstack[c]{0.649 \\ {\tiny [0.571, 0.747]}} & \shortstack[c]{0.638 \\ {\tiny [0.574, 0.737]}} \\
\shortstack[c]{\textbf{CONCH} \\ \phantom{\tiny [0.000, 0.000]}} & \shortstack[c]{0.479 \\ {\tiny [0.386, 0.554]}} & \shortstack[c]{0.713 \\ {\tiny [0.627, 0.790]}} & \shortstack[c]{0.654 \\ {\tiny [0.592, 0.743]}} & \shortstack[c]{0.561 \\ {\tiny [0.460, 0.662]}} & \shortstack[c]{0.632 \\ {\tiny [0.509, 0.703]}} & \shortstack[c]{0.633 \\ {\tiny [0.512, 0.710]}} & \shortstack[c]{0.654 \\ {\tiny [0.592, 0.743]}} & \shortstack[c]{0.651 \\ {\tiny [0.572, 0.744]}} \\
\bottomrule
\end{tabular}
}
\vspace{-8pt}
\end{table}
\vspace{-10pt}

% ================= cpm15 =================
\begin{table}[htbp]
\centering
\caption{Segmentation performance on \textbf{cpm15} under Frozen PFM. Mean with 95\% confidence interval.}
\label{tab:cpm15}
\resizebox{\textwidth}{!}{
\scriptsize
\begin{tabular}{m{2cm}cccccccc}
\toprule
\textbf{Model} & mIoU & Pixel Acc & mAcc & FW IoU & mDice & mPrecision & mRecall & mF1 \\
\midrule
\shortstack[c]{\textbf{PathOrchestra} \\ \phantom{\tiny [0.000, 0.000]}} & \shortstack[c]{0.626 \\ {\tiny [0.544, 0.721]}} & \shortstack[c]{0.883 \\ {\tiny [0.843, 0.926]}} & \shortstack[c]{0.721 \\ {\tiny [0.643, 0.808]}} & \shortstack[c]{0.806 \\ {\tiny [0.746, 0.870]}} & \shortstack[c]{0.737 \\ {\tiny [0.658, 0.822]}} & \shortstack[c]{0.758 \\ {\tiny [0.681, 0.838]}} & \shortstack[c]{0.721 \\ {\tiny [0.643, 0.808]}} & \shortstack[c]{0.737 \\ {\tiny [0.658, 0.822]}} \\
\shortstack[c]{\textbf{CONCHv1.5} \\ \phantom{\tiny [0.000, 0.000]}} & \shortstack[c]{0.634 \\ {\tiny [0.555, 0.719]}} & \shortstack[c]{0.877 \\ {\tiny [0.846, 0.910]}} & \shortstack[c]{0.762 \\ {\tiny [0.656, 0.884]}} & \shortstack[c]{0.802 \\ {\tiny [0.751, 0.854]}} & \shortstack[c]{0.748 \\ {\tiny [0.670, 0.824]}} & \shortstack[c]{0.740 \\ {\tiny [0.690, 0.787]}} & \shortstack[c]{0.762 \\ {\tiny [0.656, 0.884]}} & \shortstack[c]{0.748 \\ {\tiny [0.670, 0.824]}} \\
\shortstack[c]{\textbf{CONCH} \\ \phantom{\tiny [0.000, 0.000]}} & \shortstack[c]{0.615 \\ {\tiny [0.510, 0.721]}} & \shortstack[c]{0.881 \\ {\tiny [0.850, 0.913]}} & \shortstack[c]{0.720 \\ {\tiny [0.588, 0.871]}} & \shortstack[c]{0.801 \\ {\tiny [0.744, 0.858]}} & \shortstack[c]{0.722 \\ {\tiny [0.608, 0.825]}} & \shortstack[c]{0.746 \\ {\tiny [0.689, 0.793]}} & \shortstack[c]{0.720 \\ {\tiny [0.588, 0.871]}} & \shortstack[c]{0.722 \\ {\tiny [0.608, 0.825]}} \\
\bottomrule
\end{tabular}
}
\vspace{-8pt}
\end{table}
\vspace{-10pt}

% ================= cpm17 =================
\begin{table}[htbp]
\centering
\caption{Segmentation performance on \textbf{cpm17} under Frozen PFM. Mean with 95\% confidence interval.}
\label{tab:cpm17}
\resizebox{\textwidth}{!}{
\scriptsize
\begin{tabular}{m{2cm}cccccccc}
\toprule
\textbf{Model} & mIoU & Pixel Acc & mAcc & FW IoU & mDice & mPrecision & mRecall & mF1 \\
\midrule
\shortstack[c]{\textbf{PathOrchestra} \\ \phantom{\tiny [0.000, 0.000]}} & \shortstack[c]{0.767 \\ {\tiny [0.733, 0.797]}} & \shortstack[c]{0.919 \\ {\tiny [0.897, 0.938]}} & \shortstack[c]{0.867 \\ {\tiny [0.834, 0.898]}} & \shortstack[c]{0.857 \\ {\tiny [0.823, 0.890]}} & \shortstack[c]{0.861 \\ {\tiny [0.837, 0.882]}} & \shortstack[c]{0.856 \\ {\tiny [0.827, 0.888]}} & \shortstack[c]{0.867 \\ {\tiny [0.834, 0.898]}} & \shortstack[c]{0.861 \\ {\tiny [0.837, 0.882]}} \\
\shortstack[c]{\textbf{CONCHv1.5} \\ \phantom{\tiny [0.000, 0.000]}} & \shortstack[c]{0.764 \\ {\tiny [0.724, 0.801]}} & \shortstack[c]{0.919 \\ {\tiny [0.898, 0.941]}} & \shortstack[c]{0.856 \\ {\tiny [0.808, 0.896]}} & \shortstack[c]{0.857 \\ {\tiny [0.825, 0.893]}} & \shortstack[c]{0.858 \\ {\tiny [0.829, 0.885]}} & \shortstack[c]{0.863 \\ {\tiny [0.834, 0.900]}} & \shortstack[c]{0.856 \\ {\tiny [0.808, 0.896]}} & \shortstack[c]{0.858 \\ {\tiny [0.829, 0.885]}} \\
\shortstack[c]{\textbf{CONCH} \\ \phantom{\tiny [0.000, 0.000]}} & \shortstack[c]{0.776 \\ {\tiny [0.739, 0.808]}} & \shortstack[c]{0.925 \\ {\tiny [0.903, 0.945]}} & \shortstack[c]{0.857 \\ {\tiny [0.832, 0.879]}} & \shortstack[c]{0.866 \\ {\tiny [0.830, 0.899]}} & \shortstack[c]{0.867 \\ {\tiny [0.841, 0.889]}} & \shortstack[c]{0.877 \\ {\tiny [0.844, 0.906]}} & \shortstack[c]{0.857 \\ {\tiny [0.832, 0.879]}} & \shortstack[c]{0.867 \\ {\tiny [0.841, 0.889]}} \\
\bottomrule
\end{tabular}
}
\vspace{-8pt}
\end{table}
\vspace{-10pt}

% ================= kumar =================
\begin{table}[htbp]
\centering
\caption{Segmentation performance on \textbf{kumar} under Frozen PFM. Mean with 95\% confidence interval.}
\label{tab:kumar}
\resizebox{\textwidth}{!}{
\scriptsize
\begin{tabular}{m{2cm}cccccccc}
\toprule
\textbf{Model} & mIoU & Pixel Acc & mAcc & FW IoU & mDice & mPrecision & mRecall & mF1 \\
\midrule
\shortstack[c]{\textbf{PathOrchestra} \\ \phantom{\tiny [0.000, 0.000]}} & \shortstack[c]{0.722 \\ {\tiny [0.711, 0.727]}} & \shortstack[c]{0.886 \\ {\tiny [0.869, 0.905]}} & \shortstack[c]{0.868 \\ {\tiny [0.857, 0.874]}} & \shortstack[c]{0.811 \\ {\tiny [0.781, 0.848]}} & \shortstack[c]{0.830 \\ {\tiny [0.817, 0.835]}} & \shortstack[c]{0.806 \\ {\tiny [0.783, 0.815]}} & \shortstack[c]{0.868 \\ {\tiny [0.857, 0.874]}} & \shortstack[c]{0.830 \\ {\tiny [0.817, 0.835]}} \\
\shortstack[c]{\textbf{CONCHv1.5} \\ \phantom{\tiny [0.000, 0.000]}} & \shortstack[c]{0.729 \\ {\tiny [0.719, 0.731]}} & \shortstack[c]{0.890 \\ {\tiny [0.875, 0.917]}} & \shortstack[c]{0.869 \\ {\tiny [0.839, 0.891]}} & \shortstack[c]{0.817 \\ {\tiny [0.789, 0.860]}} & \shortstack[c]{0.836 \\ {\tiny [0.822, 0.838]}} & \shortstack[c]{0.813 \\ {\tiny [0.806, 0.822]}} & \shortstack[c]{0.869 \\ {\tiny [0.839, 0.891]}} & \shortstack[c]{0.836 \\ {\tiny [0.822, 0.838]}} \\
\shortstack[c]{\textbf{CONCH} \\ \phantom{\tiny [0.000, 0.000]}} & \shortstack[c]{0.718 \\ {\tiny [0.707, 0.727]}} & \shortstack[c]{0.880 \\ {\tiny [0.865, 0.904]}} & \shortstack[c]{0.880 \\ {\tiny [0.867, 0.890]}} & \shortstack[c]{0.804 \\ {\tiny [0.784, 0.846]}} & \shortstack[c]{0.828 \\ {\tiny [0.817, 0.837]}} & \shortstack[c]{0.799 \\ {\tiny [0.781, 0.818]}} & \shortstack[c]{0.880 \\ {\tiny [0.867, 0.890]}} & \shortstack[c]{0.828 \\ {\tiny [0.817, 0.837]}} \\
\bottomrule
\end{tabular}
}
\vspace{-8pt}
\end{table}
\vspace{-10pt}

\clearpage 
\section{LoRA Fine-tune Performance }
\renewcommand{\thefigure}{D\arabic{figure}}
\renewcommand{\thetable}{D\arabic{table}}
\setcounter{figure}{0}
\setcounter{table}{0}
% ================= BCSS =================
\begin{table}[htbp]
\centering
\caption{Segmentation performance on \textbf{BCSS} under LoRA fine-tuning. Mean with 95\% confidence interval.}
\label{tab:bcss}
\resizebox{\textwidth}{!}{
\scriptsize
\begin{tabular}{m{2cm}cccccccc}
\toprule
\textbf{Model} & mIoU & Pixel Acc & mAcc & FW IoU & mDice & mPrecision & mRecall & mF1 \\
\midrule
\shortstack[c]{\textbf{PathOrchestra} \\ \phantom{\tiny [0.000, 0.000]}} & \shortstack[c]{0.219 \\ {\tiny [0.175, 0.275]}} & \shortstack[c]{0.652 \\ {\tiny [0.569, 0.734]}} & \shortstack[c]{0.309 \\ {\tiny [0.258, 0.361]}} & \shortstack[c]{0.491 \\ {\tiny [0.394, 0.597]}} & \shortstack[c]{0.295 \\ {\tiny [0.246, 0.357]}} & \shortstack[c]{0.515 \\ {\tiny [0.436, 0.595]}} & \shortstack[c]{0.309 \\ {\tiny [0.258, 0.361]}} & \shortstack[c]{0.526 \\ {\tiny [0.434, 0.646]}} \\
\shortstack[c]{\textbf{CONCHv1.5} \\ \phantom{\tiny [0.000, 0.000]}} & \shortstack[c]{0.237 \\ {\tiny [0.198, 0.288]}} & \shortstack[c]{0.667 \\ {\tiny [0.585, 0.748]}} & \shortstack[c]{0.348 \\ {\tiny [0.314, 0.399]}} & \shortstack[c]{0.508 \\ {\tiny [0.413, 0.613]}} & \shortstack[c]{0.316 \\ {\tiny [0.270, 0.372]}} & \shortstack[c]{0.518 \\ {\tiny [0.453, 0.581]}} & \shortstack[c]{0.348 \\ {\tiny [0.314, 0.399]}} & \shortstack[c]{0.534 \\ {\tiny [0.470, 0.590]}} \\
\shortstack[c]{\textbf{CONCH} \\ \phantom{\tiny [0.000, 0.000]}} & \shortstack[c]{0.252 \\ {\tiny [0.214, 0.302]}} & \shortstack[c]{0.684 \\ {\tiny [0.603, 0.769]}} & \shortstack[c]{0.385 \\ {\tiny [0.345, 0.455]}} & \shortstack[c]{0.528 \\ {\tiny [0.426, 0.640]}} & \shortstack[c]{0.332 \\ {\tiny [0.288, 0.393]}} & \shortstack[c]{0.483 \\ {\tiny [0.424, 0.537]}} & \shortstack[c]{0.385 \\ {\tiny [0.345, 0.455]}} & \shortstack[c]{0.505 \\ {\tiny [0.445, 0.573]}} \\
\bottomrule
\end{tabular}
}
\vspace{-8pt}
\end{table}
\vspace{-10pt}

% ================= COSAS24 =================
\begin{table}[htbp]
\centering
\caption{Segmentation performance on \textbf{COSAS24} under LoRA fine-tuning. Mean with 95\% confidence interval.}
\label{tab:cosas24}
\resizebox{\textwidth}{!}{
\scriptsize
\begin{tabular}{m{2cm}cccccccc}
\toprule
\textbf{Model} & mIoU & Pixel Acc & mAcc & FW IoU & mDice & mPrecision & mRecall & mF1 \\
\midrule
\shortstack[c]{\textbf{PathOrchestra} \\ \phantom{\tiny [0.000, 0.000]}} & \shortstack[c]{0.682 \\ {\tiny [0.615, 0.750]}} & \shortstack[c]{0.814 \\ {\tiny [0.769, 0.860]}} & \shortstack[c]{0.822 \\ {\tiny [0.783, 0.862]}} & \shortstack[c]{0.692 \\ {\tiny [0.631, 0.755]}} & \shortstack[c]{0.810 \\ {\tiny [0.760, 0.857]}} & \shortstack[c]{0.809 \\ {\tiny [0.760, 0.856]}} & \shortstack[c]{0.822 \\ {\tiny [0.783, 0.862]}} & \shortstack[c]{0.810 \\ {\tiny [0.760, 0.857]}} \\
\shortstack[c]{\textbf{CONCHv1.5} \\ \phantom{\tiny [0.000, 0.000]}} & \shortstack[c]{0.766 \\ {\tiny [0.703, 0.826]}} & \shortstack[c]{0.870 \\ {\tiny [0.831, 0.906]}} & \shortstack[c]{0.879 \\ {\tiny [0.848, 0.909]}} & \shortstack[c]{0.774 \\ {\tiny [0.718, 0.829]}} & \shortstack[c]{0.867 \\ {\tiny [0.824, 0.905]}} & \shortstack[c]{0.864 \\ {\tiny [0.817, 0.903]}} & \shortstack[c]{0.879 \\ {\tiny [0.848, 0.909]}} & \shortstack[c]{0.867 \\ {\tiny [0.824, 0.905]}} \\
\shortstack[c]{\textbf{CONCH} \\ \phantom{\tiny [0.000, 0.000]}} & \shortstack[c]{0.729 \\ {\tiny [0.668, 0.791]}} & \shortstack[c]{0.847 \\ {\tiny [0.807, 0.885]}} & \shortstack[c]{0.854 \\ {\tiny [0.813, 0.888]}} & \shortstack[c]{0.738 \\ {\tiny [0.680, 0.795]}} & \shortstack[c]{0.842 \\ {\tiny [0.799, 0.883]}} & \shortstack[c]{0.840 \\ {\tiny [0.794, 0.882]}} & \shortstack[c]{0.854 \\ {\tiny [0.813, 0.888]}} & \shortstack[c]{0.842 \\ {\tiny [0.799, 0.883]}} \\
\bottomrule
\end{tabular}
}
\vspace{-8pt}
\end{table}
\vspace{-10pt}

% ================= CRAG =================
\begin{table}[htbp]
\centering
\caption{Segmentation performance on \textbf{CRAG} under LoRA fine-tuning. Mean with 95\% confidence interval.}
\label{tab:crag}
\resizebox{\textwidth}{!}{
\scriptsize
\begin{tabular}{m{2cm}cccccccc}
\toprule
\textbf{Model} & mIoU & Pixel Acc & mAcc & FW IoU & mDice & mPrecision & mRecall & mF1 \\
\midrule
\shortstack[c]{\textbf{PathOrchestra} \\ \phantom{\tiny [0.000, 0.000]}} & \shortstack[c]{0.826 \\ {\tiny [0.773, 0.872]}} & \shortstack[c]{0.910 \\ {\tiny [0.875, 0.939]}} & \shortstack[c]{0.916 \\ {\tiny [0.888, 0.940]}} & \shortstack[c]{0.838 \\ {\tiny [0.781, 0.887]}} & \shortstack[c]{0.904 \\ {\tiny [0.872, 0.931]}} & \shortstack[c]{0.897 \\ {\tiny [0.865, 0.926]}} & \shortstack[c]{0.916 \\ {\tiny [0.888, 0.940]}} & \shortstack[c]{0.904 \\ {\tiny [0.872, 0.931]}} \\
\shortstack[c]{\textbf{CONCHv1.5} \\ \phantom{\tiny [0.000, 0.000]}} & \shortstack[c]{0.858 \\ {\tiny [0.797, 0.907]}} & \shortstack[c]{0.928 \\ {\tiny [0.893, 0.957]}} & \shortstack[c]{0.931 \\ {\tiny [0.901, 0.954]}} & \shortstack[c]{0.868 \\ {\tiny [0.809, 0.918]}} & \shortstack[c]{0.923 \\ {\tiny [0.887, 0.951]}} & \shortstack[c]{0.917 \\ {\tiny [0.879, 0.949]}} & \shortstack[c]{0.931 \\ {\tiny [0.901, 0.954]}} & \shortstack[c]{0.923 \\ {\tiny [0.887, 0.951]}} \\
\shortstack[c]{\textbf{CONCH} \\ \phantom{\tiny [0.000, 0.000]}} & \shortstack[c]{0.853 \\ {\tiny [0.795, 0.901]}} & \shortstack[c]{0.926 \\ {\tiny [0.891, 0.955]}} & \shortstack[c]{0.926 \\ {\tiny [0.898, 0.948]}} & \shortstack[c]{0.864 \\ {\tiny [0.806, 0.913]}} & \shortstack[c]{0.920 \\ {\tiny [0.885, 0.948]}} & \shortstack[c]{0.916 \\ {\tiny [0.879, 0.948]}} & \shortstack[c]{0.926 \\ {\tiny [0.898, 0.948]}} & \shortstack[c]{0.920 \\ {\tiny [0.885, 0.948]}} \\
\bottomrule
\end{tabular}
}
\vspace{-8pt}
\end{table}
\vspace{-10pt}

% ================= CoCaHis =================
\begin{table}[htbp]
\centering
\caption{Segmentation performance on \textbf{CoCaHis} under LoRA fine-tuning. Mean with 95\% confidence interval.}
\label{tab:cocahis}
\resizebox{\textwidth}{!}{
\scriptsize
\begin{tabular}{m{2cm}cccccccc}
\toprule
\textbf{Model} & mIoU & Pixel Acc & mAcc & FW IoU & mDice & mPrecision & mRecall & mF1 \\
\midrule
\shortstack[c]{\textbf{PathOrchestra} \\ \phantom{\tiny [0.000, 0.000]}} & \shortstack[c]{0.819 \\ {\tiny [0.788, 0.848]}} & \shortstack[c]{0.906 \\ {\tiny [0.882, 0.929]}} & \shortstack[c]{0.906 \\ {\tiny [0.884, 0.927]}} & \shortstack[c]{0.831 \\ {\tiny [0.790, 0.870]}} & \shortstack[c]{0.900 \\ {\tiny [0.881, 0.917]}} & \shortstack[c]{0.896 \\ {\tiny [0.880, 0.911]}} & \shortstack[c]{0.906 \\ {\tiny [0.884, 0.927]}} & \shortstack[c]{0.900 \\ {\tiny [0.881, 0.917]}} \\
\shortstack[c]{\textbf{CONCHv1.5} \\ \phantom{\tiny [0.000, 0.000]}} & \shortstack[c]{0.859 \\ {\tiny [0.831, 0.885]}} & \shortstack[c]{0.929 \\ {\tiny [0.908, 0.948]}} & \shortstack[c]{0.928 \\ {\tiny [0.908, 0.944]}} & \shortstack[c]{0.868 \\ {\tiny [0.832, 0.903]}} & \shortstack[c]{0.924 \\ {\tiny [0.908, 0.939]}} & \shortstack[c]{0.921 \\ {\tiny [0.906, 0.935]}} & \shortstack[c]{0.928 \\ {\tiny [0.908, 0.944]}} & \shortstack[c]{0.924 \\ {\tiny [0.908, 0.939]}} \\
\shortstack[c]{\textbf{CONCH} \\ \phantom{\tiny [0.000, 0.000]}} & \shortstack[c]{0.854 \\ {\tiny [0.825, 0.881]}} & \shortstack[c]{0.926 \\ {\tiny [0.905, 0.946]}} & \shortstack[c]{0.925 \\ {\tiny [0.906, 0.941]}} & \shortstack[c]{0.864 \\ {\tiny [0.826, 0.899]}} & \shortstack[c]{0.921 \\ {\tiny [0.904, 0.936]}} & \shortstack[c]{0.918 \\ {\tiny [0.904, 0.932]}} & \shortstack[c]{0.925 \\ {\tiny [0.906, 0.941]}} & \shortstack[c]{0.921 \\ {\tiny [0.904, 0.936]}} \\
\bottomrule
\end{tabular}
}
\vspace{-8pt}
\end{table}
\vspace{-10pt}

% ================= CoNIC2022 =================
\begin{table}[htbp]
\centering
\caption{Segmentation performance on \textbf{CoNIC2022} under LoRA fine-tuning. Mean with 95\% confidence interval.}
\label{tab:conic2022}
\resizebox{\textwidth}{!}{
\scriptsize
\begin{tabular}{m{2cm}cccccccc}
\toprule
\textbf{Model} & mIoU & Pixel Acc & mAcc & FW IoU & mDice & mPrecision & mRecall & mF1 \\
\midrule
\shortstack[c]{\textbf{PathOrchestra} \\ \phantom{\tiny [0.000, 0.000]}} & \shortstack[c]{0.359 \\ {\tiny [0.352, 0.367]}} & \shortstack[c]{0.874 \\ {\tiny [0.868, 0.879]}} & \shortstack[c]{0.561 \\ {\tiny [0.548, 0.575]}} & \shortstack[c]{0.802 \\ {\tiny [0.793, 0.811]}} & \shortstack[c]{0.489 \\ {\tiny [0.479, 0.500]}} & \shortstack[c]{0.447 \\ {\tiny [0.438, 0.457]}} & \shortstack[c]{0.561 \\ {\tiny [0.548, 0.575]}} & \shortstack[c]{0.489 \\ {\tiny [0.479, 0.500]}} \\
\shortstack[c]{\textbf{CONCHv1.5} \\ \phantom{\tiny [0.000, 0.000]}} & \shortstack[c]{0.407 \\ {\tiny [0.399, 0.415]}} & \shortstack[c]{0.895 \\ {\tiny [0.890, 0.900]}} & \shortstack[c]{0.594 \\ {\tiny [0.583, 0.605]}} & \shortstack[c]{0.830 \\ {\tiny [0.822, 0.837]}} & \shortstack[c]{0.546 \\ {\tiny [0.535, 0.556]}} & \shortstack[c]{0.517 \\ {\tiny [0.506, 0.528]}} & \shortstack[c]{0.594 \\ {\tiny [0.583, 0.605]}} & \shortstack[c]{0.546 \\ {\tiny [0.535, 0.556]}} \\
\shortstack[c]{\textbf{CONCH} \\ \phantom{\tiny [0.000, 0.000]}} & \shortstack[c]{0.413 \\ {\tiny [0.405, 0.420]}} & \shortstack[c]{0.897 \\ {\tiny [0.893, 0.902]}} & \shortstack[c]{0.591 \\ {\tiny [0.580, 0.602]}} & \shortstack[c]{0.831 \\ {\tiny [0.824, 0.839]}} & \shortstack[c]{0.553 \\ {\tiny [0.542, 0.563]}} & \shortstack[c]{0.530 \\ {\tiny [0.519, 0.541]}} & \shortstack[c]{0.591 \\ {\tiny [0.580, 0.602]}} & \shortstack[c]{0.553 \\ {\tiny [0.542, 0.563]}} \\
\bottomrule
\end{tabular}
}
\vspace{-8pt}
\end{table}
\vspace{-10pt}

% ================= CoNSeP =================
\begin{table}[htbp]
\centering
\caption{Segmentation performance on \textbf{CoNSeP} under LoRA fine-tuning. Mean with 95\% confidence interval.}
\label{tab:consep}
\resizebox{\textwidth}{!}{
\scriptsize
\begin{tabular}{m{2cm}cccccccc}
\toprule
\textbf{Model} & mIoU & Pixel Acc & mAcc & FW IoU & mDice & mPrecision & mRecall & mF1 \\
\midrule
\shortstack[c]{\textbf{PathOrchestra} \\ \phantom{\tiny [0.000, 0.000]}} & \shortstack[c]{0.263 \\ {\tiny [0.195, 0.385]}} & \shortstack[c]{0.868 \\ {\tiny [0.828, 0.909]}} & \shortstack[c]{0.453 \\ {\tiny [0.366, 0.604]}} & \shortstack[c]{0.822 \\ {\tiny [0.770, 0.879]}} & \shortstack[c]{0.347 \\ {\tiny [0.250, 0.509]}} & \shortstack[c]{0.392 \\ {\tiny [0.333, 0.468]}} & \shortstack[c]{0.453 \\ {\tiny [0.366, 0.604]}} & \shortstack[c]{0.451 \\ {\tiny [0.392, 0.564]}} \\
\shortstack[c]{\textbf{CONCHv1.5} \\ \phantom{\tiny [0.000, 0.000]}} & \shortstack[c]{0.316 \\ {\tiny [0.239, 0.456]}} & \shortstack[c]{0.896 \\ {\tiny [0.847, 0.943]}} & \shortstack[c]{0.490 \\ {\tiny [0.383, 0.641]}} & \shortstack[c]{0.850 \\ {\tiny [0.785, 0.913]}} & \shortstack[c]{0.405 \\ {\tiny [0.305, 0.573]}} & \shortstack[c]{0.461 \\ {\tiny [0.386, 0.571]}} & \shortstack[c]{0.490 \\ {\tiny [0.383, 0.641]}} & \shortstack[c]{0.629 \\ {\tiny [0.569, 0.688]}} \\
\shortstack[c]{\textbf{CONCH} \\ \phantom{\tiny [0.000, 0.000]}} & \shortstack[c]{0.316 \\ {\tiny [0.239, 0.392]}} & \shortstack[c]{0.883 \\ {\tiny [0.845, 0.921]}} & \shortstack[c]{0.552 \\ {\tiny [0.434, 0.699]}} & \shortstack[c]{0.835 \\ {\tiny [0.786, 0.887]}} & \shortstack[c]{0.419 \\ {\tiny [0.305, 0.531]}} & \shortstack[c]{0.437 \\ {\tiny [0.341, 0.497]}} & \shortstack[c]{0.552 \\ {\tiny [0.434, 0.699]}} & \shortstack[c]{0.536 \\ {\tiny [0.468, 0.613]}} \\
\bottomrule
\end{tabular}
}
\vspace{-8pt}
\end{table}
\vspace{-10pt}

% ================= EBHI =================
\begin{table}[htbp]
\centering
\caption{Segmentation performance on \textbf{EBHI} under LoRA fine-tuning. Mean with 95\% confidence interval.}
\label{tab:ebhi}
\resizebox{\textwidth}{!}{
\scriptsize
\begin{tabular}{m{2cm}cccccccc}
\toprule
\textbf{Model} & mIoU & Pixel Acc & mAcc & FW IoU & mDice & mPrecision & mRecall & mF1 \\
\midrule
\shortstack[c]{\textbf{PathOrchestra} \\ \phantom{\tiny [0.000, 0.000]}} & \shortstack[c]{0.427 \\ {\tiny [0.405, 0.446]}} & \shortstack[c]{0.821 \\ {\tiny [0.787, 0.851]}} & \shortstack[c]{0.523 \\ {\tiny [0.517, 0.527]}} & \shortstack[c]{0.690 \\ {\tiny [0.641, 0.736]}} & \shortstack[c]{0.488 \\ {\tiny [0.472, 0.501]}} & \shortstack[c]{0.804 \\ {\tiny [0.764, 0.840]}} & \shortstack[c]{0.523 \\ {\tiny [0.517, 0.527]}} & \shortstack[c]{0.853 \\ {\tiny [0.826, 0.875]}} \\
\shortstack[c]{\textbf{CONCHv1.5} \\ \phantom{\tiny [0.000, 0.000]}} & \shortstack[c]{0.436 \\ {\tiny [0.413, 0.455]}} & \shortstack[c]{0.830 \\ {\tiny [0.796, 0.860]}} & \shortstack[c]{0.530 \\ {\tiny [0.521, 0.536]}} & \shortstack[c]{0.703 \\ {\tiny [0.654, 0.750]}} & \shortstack[c]{0.494 \\ {\tiny [0.478, 0.507]}} & \shortstack[c]{0.813 \\ {\tiny [0.773, 0.850]}} & \shortstack[c]{0.530 \\ {\tiny [0.521, 0.536]}} & \shortstack[c]{0.863 \\ {\tiny [0.836, 0.886]}} \\
\shortstack[c]{\textbf{CONCH} \\ \phantom{\tiny [0.000, 0.000]}} & \shortstack[c]{0.433 \\ {\tiny [0.409, 0.454]}} & \shortstack[c]{0.827 \\ {\tiny [0.794, 0.858]}} & \shortstack[c]{0.527 \\ {\tiny [0.517, 0.535]}} & \shortstack[c]{0.699 \\ {\tiny [0.652, 0.748]}} & \shortstack[c]{0.492 \\ {\tiny [0.475, 0.506]}} & \shortstack[c]{0.812 \\ {\tiny [0.771, 0.849]}} & \shortstack[c]{0.527 \\ {\tiny [0.517, 0.535]}} & \shortstack[c]{0.860 \\ {\tiny [0.831, 0.884]}} \\
\bottomrule
\end{tabular}
}
\vspace{-8pt}
\end{table}
\vspace{-10pt}

% ================= GlaS =================
\begin{table}[htbp]
\centering
\caption{Segmentation performance on \textbf{GlaS} under LoRA fine-tuning. Mean with 95\% confidence interval.}
\label{tab:glas}
\resizebox{\textwidth}{!}{
\scriptsize
\begin{tabular}{m{2cm}cccccccc}
\toprule
\textbf{Model} & mIoU & Pixel Acc & mAcc & FW IoU & mDice & mPrecision & mRecall & mF1 \\
\midrule
\shortstack[c]{\textbf{PathOrchestra} \\ \phantom{\tiny [0.000, 0.000]}} & \shortstack[c]{0.878 \\ {\tiny [0.854, 0.899]}} & \shortstack[c]{0.935 \\ {\tiny [0.922, 0.948]}} & \shortstack[c]{0.935 \\ {\tiny [0.921, 0.948]}} & \shortstack[c]{0.879 \\ {\tiny [0.855, 0.901]}} & \shortstack[c]{0.935 \\ {\tiny [0.921, 0.947]}} & \shortstack[c]{0.935 \\ {\tiny [0.922, 0.946]}} & \shortstack[c]{0.935 \\ {\tiny [0.921, 0.948]}} & \shortstack[c]{0.935 \\ {\tiny [0.921, 0.947]}} \\
\shortstack[c]{\textbf{CONCHv1.5} \\ \phantom{\tiny [0.000, 0.000]}} & \shortstack[c]{0.893 \\ {\tiny [0.874, 0.910]}} & \shortstack[c]{0.944 \\ {\tiny [0.933, 0.954]}} & \shortstack[c]{0.944 \\ {\tiny [0.933, 0.953]}} & \shortstack[c]{0.894 \\ {\tiny [0.874, 0.913]}} & \shortstack[c]{0.944 \\ {\tiny [0.933, 0.953]}} & \shortstack[c]{0.944 \\ {\tiny [0.933, 0.953]}} & \shortstack[c]{0.944 \\ {\tiny [0.933, 0.953]}} & \shortstack[c]{0.944 \\ {\tiny [0.933, 0.953]}} \\
\shortstack[c]{\textbf{CONCH} \\ \phantom{\tiny [0.000, 0.000]}} & \shortstack[c]{0.884 \\ {\tiny [0.864, 0.903]}} & \shortstack[c]{0.939 \\ {\tiny [0.927, 0.950]}} & \shortstack[c]{0.939 \\ {\tiny [0.927, 0.949]}} & \shortstack[c]{0.885 \\ {\tiny [0.864, 0.906]}} & \shortstack[c]{0.939 \\ {\tiny [0.927, 0.949]}} & \shortstack[c]{0.938 \\ {\tiny [0.927, 0.949]}} & \shortstack[c]{0.939 \\ {\tiny [0.927, 0.949]}} & \shortstack[c]{0.939 \\ {\tiny [0.927, 0.949]}} \\
\bottomrule
\end{tabular}
}
\vspace{-8pt}
\end{table}
\vspace{-10pt}

% ================= Janowczyk =================
\begin{table}[htbp]
\centering
\caption{Segmentation performance on \textbf{Janowczyk} under LoRA fine-tuning. Mean with 95\% confidence interval.}
\label{tab:janowczyk}
\resizebox{\textwidth}{!}{
\scriptsize
\begin{tabular}{m{2cm}cccccccc}
\toprule
\textbf{Model} & mIoU & Pixel Acc & mAcc & FW IoU & mDice & mPrecision & mRecall & mF1 \\
\midrule
\shortstack[c]{\textbf{PathOrchestra} \\ \phantom{\tiny [0.000, 0.000]}} & \shortstack[c]{0.573 \\ {\tiny [0.549, 0.596]}} & \shortstack[c]{0.973 \\ {\tiny [0.968, 0.978]}} & \shortstack[c]{0.632 \\ {\tiny [0.588, 0.675]}} & \shortstack[c]{0.957 \\ {\tiny [0.948, 0.965]}} & \shortstack[c]{0.640 \\ {\tiny [0.603, 0.672]}} & \shortstack[c]{0.651 \\ {\tiny [0.614, 0.692]}} & \shortstack[c]{0.632 \\ {\tiny [0.588, 0.675]}} & \shortstack[c]{0.640 \\ {\tiny [0.603, 0.672]}} \\
\shortstack[c]{\textbf{CONCHv1.5} \\ \phantom{\tiny [0.000, 0.000]}} & \shortstack[c]{0.595 \\ {\tiny [0.578, 0.618]}} & \shortstack[c]{0.973 \\ {\tiny [0.969, 0.977]}} & \shortstack[c]{0.677 \\ {\tiny [0.646, 0.714]}} & \shortstack[c]{0.958 \\ {\tiny [0.949, 0.965]}} & \shortstack[c]{0.672 \\ {\tiny [0.647, 0.701]}} & \shortstack[c]{0.667 \\ {\tiny [0.636, 0.706]}} & \shortstack[c]{0.677 \\ {\tiny [0.646, 0.714]}} & \shortstack[c]{0.672 \\ {\tiny [0.647, 0.701]}} \\
\shortstack[c]{\textbf{CONCH} \\ \phantom{\tiny [0.000, 0.000]}} & \shortstack[c]{0.594 \\ {\tiny [0.574, 0.618]}} & \shortstack[c]{0.971 \\ {\tiny [0.966, 0.976]}} & \shortstack[c]{0.687 \\ {\tiny [0.643, 0.733]}} & \shortstack[c]{0.956 \\ {\tiny [0.947, 0.964]}} & \shortstack[c]{0.670 \\ {\tiny [0.643, 0.702]}} & \shortstack[c]{0.658 \\ {\tiny [0.628, 0.694]}} & \shortstack[c]{0.687 \\ {\tiny [0.643, 0.733]}} & \shortstack[c]{0.670 \\ {\tiny [0.643, 0.702]}} \\
\bottomrule
\end{tabular}
}
\vspace{-8pt}
\end{table}
\vspace{-10pt}

% ================= Lizard =================
\begin{table}[htbp]
\centering
\caption{Segmentation performance on \textbf{Lizard} under LoRA fine-tuning. Mean with 95\% confidence interval.}
\label{tab:lizard}
\resizebox{\textwidth}{!}{
\scriptsize
\begin{tabular}{m{2cm}cccccccc}
\toprule
\textbf{Model} & mIoU & Pixel Acc & mAcc & FW IoU & mDice & mPrecision & mRecall & mF1 \\
\midrule
\shortstack[c]{\textbf{PathOrchestra} \\ \phantom{\tiny [0.000, 0.000]}} & \shortstack[c]{0.213 \\ {\tiny [0.190, 0.233]}} & \shortstack[c]{0.793 \\ {\tiny [0.765, 0.821]}} & \shortstack[c]{0.317 \\ {\tiny [0.275, 0.354]}} & \shortstack[c]{0.700 \\ {\tiny [0.662, 0.738]}} & \shortstack[c]{0.288 \\ {\tiny [0.254, 0.317]}} & \shortstack[c]{0.379 \\ {\tiny [0.338, 0.415]}} & \shortstack[c]{0.317 \\ {\tiny [0.275, 0.354]}} & \shortstack[c]{0.404 \\ {\tiny [0.356, 0.444]}} \\
\shortstack[c]{\textbf{CONCHv1.5} \\ \phantom{\tiny [0.000, 0.000]}} & \shortstack[c]{0.228 \\ {\tiny [0.205, 0.249]}} & \shortstack[c]{0.804 \\ {\tiny [0.780, 0.830]}} & \shortstack[c]{0.337 \\ {\tiny [0.301, 0.367]}} & \shortstack[c]{0.713 \\ {\tiny [0.677, 0.750]}} & \shortstack[c]{0.308 \\ {\tiny [0.273, 0.337]}} & \shortstack[c]{0.338 \\ {\tiny [0.303, 0.369]}} & \shortstack[c]{0.337 \\ {\tiny [0.301, 0.367]}} & \shortstack[c]{0.431 \\ {\tiny [0.382, 0.472]}} \\
\shortstack[c]{\textbf{CONCH} \\ \phantom{\tiny [0.000, 0.000]}} & \shortstack[c]{0.223 \\ {\tiny [0.201, 0.244]}} & \shortstack[c]{0.797 \\ {\tiny [0.770, 0.824]}} & \shortstack[c]{0.330 \\ {\tiny [0.297, 0.362]}} & \shortstack[c]{0.705 \\ {\tiny [0.668, 0.743]}} & \shortstack[c]{0.302 \\ {\tiny [0.269, 0.331]}} & \shortstack[c]{0.331 \\ {\tiny [0.295, 0.363]}} & \shortstack[c]{0.330 \\ {\tiny [0.297, 0.362]}} & \shortstack[c]{0.354 \\ {\tiny [0.316, 0.416]}} \\
\bottomrule
\end{tabular}
}
\vspace{-8pt}
\end{table}
\vspace{-10pt}

% ================= NuCLS =================
\begin{table}[htbp]
\centering
\caption{Segmentation performance on \textbf{NuCLS} under LoRA fine-tuning. Mean with 95\% confidence interval.}
\label{tab:nucls}
\resizebox{\textwidth}{!}{
\scriptsize
\begin{tabular}{m{2cm}cccccccc}
\toprule
\textbf{Model} & mIoU & Pixel Acc & mAcc & FW IoU & mDice & mPrecision & mRecall & mF1 \\
\midrule
\shortstack[c]{\textbf{PathOrchestra} \\ \phantom{\tiny [0.000, 0.000]}} & \shortstack[c]{0.212 \\ {\tiny [0.195, 0.252]}} & \shortstack[c]{0.788 \\ {\tiny [0.777, 0.798]}} & \shortstack[c]{0.311 \\ {\tiny [0.285, 0.369]}} & \shortstack[c]{0.680 \\ {\tiny [0.666, 0.695]}} & \shortstack[c]{0.286 \\ {\tiny [0.263, 0.339]}} & \shortstack[c]{0.472 \\ {\tiny [0.453, 0.492]}} & \shortstack[c]{0.311 \\ {\tiny [0.285, 0.369]}} & \shortstack[c]{0.499 \\ {\tiny [0.479, 0.519]}} \\
\shortstack[c]{\textbf{CONCHv1.5} \\ \phantom{\tiny [0.000, 0.000]}} & \shortstack[c]{0.223 \\ {\tiny [0.205, 0.265]}} & \shortstack[c]{0.802 \\ {\tiny [0.791, 0.812]}} & \shortstack[c]{0.325 \\ {\tiny [0.296, 0.385]}} & \shortstack[c]{0.700 \\ {\tiny [0.686, 0.714]}} & \shortstack[c]{0.300 \\ {\tiny [0.275, 0.355]}} & \shortstack[c]{0.498 \\ {\tiny [0.475, 0.520]}} & \shortstack[c]{0.325 \\ {\tiny [0.296, 0.385]}} & \shortstack[c]{0.524 \\ {\tiny [0.501, 0.546]}} \\
\shortstack[c]{\textbf{CONCH} \\ \phantom{\tiny [0.000, 0.000]}} & \shortstack[c]{0.223 \\ {\tiny [0.205, 0.264]}} & \shortstack[c]{0.801 \\ {\tiny [0.790, 0.811]}} & \shortstack[c]{0.319 \\ {\tiny [0.294, 0.378]}} & \shortstack[c]{0.696 \\ {\tiny [0.682, 0.710]}} & \shortstack[c]{0.299 \\ {\tiny [0.275, 0.354]}} & \shortstack[c]{0.498 \\ {\tiny [0.476, 0.521]}} & \shortstack[c]{0.319 \\ {\tiny [0.294, 0.378]}} & \shortstack[c]{0.522 \\ {\tiny [0.502, 0.542]}} \\
\bottomrule
\end{tabular}
}
\vspace{-8pt}
\end{table}
\vspace{-10pt}

% ================= PanNuke =================
\begin{table}[htbp]
\centering
\caption{Segmentation performance on \textbf{PanNuke} under LoRA fine-tuning. Mean with 95\% confidence interval.}
\label{tab:pannuke}
\resizebox{\textwidth}{!}{
\scriptsize
\begin{tabular}{m{2cm}cccccccc}
\toprule
\textbf{Model} & mIoU & Pixel Acc & mAcc & FW IoU & mDice & mPrecision & mRecall & mF1 \\
\midrule
\shortstack[c]{\textbf{PathOrchestra} \\ \phantom{\tiny [0.000, 0.000]}} & \shortstack[c]{0.471 \\ {\tiny [0.461, 0.481]}} & \shortstack[c]{0.898 \\ {\tiny [0.891, 0.904]}} & \shortstack[c]{0.612 \\ {\tiny [0.602, 0.621]}} & \shortstack[c]{0.832 \\ {\tiny [0.820, 0.842]}} & \shortstack[c]{0.588 \\ {\tiny [0.579, 0.597]}} & \shortstack[c]{0.684 \\ {\tiny [0.671, 0.696]}} & \shortstack[c]{0.612 \\ {\tiny [0.602, 0.621]}} & \shortstack[c]{0.706 \\ {\tiny [0.695, 0.717]}} \\
\shortstack[c]{\textbf{CONCHv1.5} \\ \phantom{\tiny [0.000, 0.000]}} & \shortstack[c]{0.517 \\ {\tiny [0.507, 0.526]}} & \shortstack[c]{0.920 \\ {\tiny [0.916, 0.923]}} & \shortstack[c]{0.639 \\ {\tiny [0.632, 0.646]}} & \shortstack[c]{0.864 \\ {\tiny [0.858, 0.870]}} & \shortstack[c]{0.625 \\ {\tiny [0.617, 0.634]}} & \shortstack[c]{0.736 \\ {\tiny [0.724, 0.747]}} & \shortstack[c]{0.639 \\ {\tiny [0.632, 0.646]}} & \shortstack[c]{0.750 \\ {\tiny [0.740, 0.760]}} \\
\shortstack[c]{\textbf{CONCH} \\ \phantom{\tiny [0.000, 0.000]}} & \shortstack[c]{0.518 \\ {\tiny [0.508, 0.527]}} & \shortstack[c]{0.922 \\ {\tiny [0.919, 0.926]}} & \shortstack[c]{0.630 \\ {\tiny [0.621, 0.637]}} & \shortstack[c]{0.867 \\ {\tiny [0.862, 0.873]}} & \shortstack[c]{0.626 \\ {\tiny [0.617, 0.634]}} & \shortstack[c]{0.748 \\ {\tiny [0.737, 0.759]}} & \shortstack[c]{0.630 \\ {\tiny [0.621, 0.637]}} & \shortstack[c]{0.751 \\ {\tiny [0.741, 0.761]}} \\
\bottomrule
\end{tabular}
}
\vspace{-8pt}
\end{table}
\vspace{-10pt}

% ================= RINGS =================
\begin{table}[htbp]
\centering
\caption{Segmentation performance on \textbf{RINGS} under LoRA fine-tuning. Mean with 95\% confidence interval.}
\label{tab:rings}
\resizebox{\textwidth}{!}{
\scriptsize
\begin{tabular}{m{2cm}cccccccc}
\toprule
\textbf{Model} & mIoU & Pixel Acc & mAcc & FW IoU & mDice & mPrecision & mRecall & mF1 \\
\midrule
\shortstack[c]{\textbf{PathOrchestra} \\ \phantom{\tiny [0.000, 0.000]}} & \shortstack[c]{0.876 \\ {\tiny [0.862, 0.890]}} & \shortstack[c]{0.945 \\ {\tiny [0.937, 0.952]}} & \shortstack[c]{0.933 \\ {\tiny [0.924, 0.942]}} & \shortstack[c]{0.896 \\ {\tiny [0.882, 0.908]}} & \shortstack[c]{0.934 \\ {\tiny [0.925, 0.942]}} & \shortstack[c]{0.934 \\ {\tiny [0.925, 0.943]}} & \shortstack[c]{0.933 \\ {\tiny [0.924, 0.942]}} & \shortstack[c]{0.934 \\ {\tiny [0.925, 0.942]}} \\
\shortstack[c]{\textbf{CONCHv1.5} \\ \phantom{\tiny [0.000, 0.000]}} & \shortstack[c]{0.894 \\ {\tiny [0.877, 0.909]}} & \shortstack[c]{0.953 \\ {\tiny [0.946, 0.959]}} & \shortstack[c]{0.947 \\ {\tiny [0.939, 0.956]}} & \shortstack[c]{0.910 \\ {\tiny [0.898, 0.923]}} & \shortstack[c]{0.943 \\ {\tiny [0.934, 0.952]}} & \shortstack[c]{0.940 \\ {\tiny [0.929, 0.949]}} & \shortstack[c]{0.947 \\ {\tiny [0.939, 0.956]}} & \shortstack[c]{0.943 \\ {\tiny [0.934, 0.952]}} \\
\shortstack[c]{\textbf{CONCH} \\ \phantom{\tiny [0.000, 0.000]}} & \shortstack[c]{0.893 \\ {\tiny [0.879, 0.908]}} & \shortstack[c]{0.952 \\ {\tiny [0.946, 0.959]}} & \shortstack[c]{0.949 \\ {\tiny [0.942, 0.956]}} & \shortstack[c]{0.910 \\ {\tiny [0.898, 0.921]}} & \shortstack[c]{0.943 \\ {\tiny [0.935, 0.951]}} & \shortstack[c]{0.938 \\ {\tiny [0.928, 0.947]}} & \shortstack[c]{0.949 \\ {\tiny [0.942, 0.956]}} & \shortstack[c]{0.943 \\ {\tiny [0.935, 0.951]}} \\
\bottomrule
\end{tabular}
}
\vspace{-8pt}
\end{table}
\vspace{-10pt}

% ================= TNBC =================
\begin{table}[htbp]
\centering
\caption{Segmentation performance on \textbf{TNBC} under LoRA fine-tuning. Mean with 95\% confidence interval.}
\label{tab:tnbc}
\resizebox{\textwidth}{!}{
\scriptsize
%
}
\vspace{-8pt}
\end{table}
\vspace{-10pt}

% ================= WSSS4LUAD =================
\begin{table}[htbp]
\centering
\caption{Segmentation performance on \textbf{WSSS4LUAD} under LoRA fine-tuning. Mean with 95\% confidence interval.}
\label{tab:wsss4luad}
\resizebox{\textwidth}{!}{
\scriptsize
%
}
\vspace{-8pt}
\end{table}
\vspace{-10pt}

% ================= cpm15 =================
\begin{table}[htbp]
\centering
\caption{Segmentation performance on \textbf{cpm15} under LoRA fine-tuning. Mean with 95\% confidence interval.}
\label{tab:cpm15}
\resizebox{\textwidth}{!}{
\scriptsize
%
}
\vspace{-8pt}
\end{table}
\vspace{-10pt}

% ================= cpm17 =================
\begin{table}[htbp]
\centering
\caption{Segmentation performance on \textbf{cpm17} under LoRA fine-tuning. Mean with 95\% confidence interval.}
\label{tab:cpm17}
\resizebox{\textwidth}{!}{
\scriptsize
%
}
\vspace{-8pt}
\end{table}
\vspace{-10pt}

% ================= kumar =================
\begin{table}[htbp]
\centering
\caption{Segmentation performance on \textbf{kumar} under LoRA fine-tuning. Mean with 95\% confidence interval.}
\label{tab:kumar}
\resizebox{\textwidth}{!}{
\scriptsize
%
}
\vspace{-8pt}
\end{table}
\vspace{-10pt}

\clearpage 
\section{DoRA Fine-tune Performance }
\renewcommand{\thefigure}{E\arabic{figure}}
\renewcommand{\thetable}{E\arabic{table}}
\setcounter{figure}{0}
\setcounter{table}{0}
% ================= BCSS =================
\begin{table}[htbp]
\centering
\caption{Segmentation performance on \textbf{BCSS} under DoRA fine-tuning. Mean with 95\% confidence interval.}
\label{tab:bcss}
\resizebox{\textwidth}{!}{
\scriptsize
%
}
\vspace{-8pt}
\end{table}
\vspace{-10pt}

% ================= COSAS24 =================
\begin{table}[htbp]
\centering
\caption{Segmentation performance on \textbf{COSAS24} under DoRA fine-tuning. Mean with 95\% confidence interval.}
\label{tab:cosas24}
\resizebox{\textwidth}{!}{
\scriptsize
%
}
\vspace{-8pt}
\end{table}
\vspace{-10pt}

% ================= CRAG =================
\begin{table}[htbp]
\centering
\caption{Segmentation performance on \textbf{CRAG} under DoRA fine-tuning. Mean with 95\% confidence interval.}
\label{tab:crag}
\resizebox{\textwidth}{!}{
\scriptsize
%
}
\vspace{-8pt}
\end{table}
\vspace{-10pt}

% ================= CoCaHis =================
\begin{table}[htbp]
\centering
\caption{Segmentation performance on \textbf{CoCaHis} under DoRA fine-tuning. Mean with 95\% confidence interval.}
\label{tab:cocahis}
\resizebox{\textwidth}{!}{
\scriptsize
%
}
\vspace{-8pt}
\end{table}
\vspace{-10pt}

% ================= CoNIC2022 =================
\begin{table}[htbp]
\centering
\caption{Segmentation performance on \textbf{CoNIC2022} under DoRA fine-tuning. Mean with 95\% confidence interval.}
\label{tab:conic2022}
\resizebox{\textwidth}{!}{
\scriptsize
%
}
\vspace{-8pt}
\end{table}
\vspace{-10pt}

% ================= CoNSeP =================
\begin{table}[htbp]
\centering
\caption{Segmentation performance on \textbf{CoNSeP} under DoRA fine-tuning. Mean with 95\% confidence interval.}
\label{tab:consep}
\resizebox{\textwidth}{!}{
\scriptsize
%
}
\vspace{-8pt}
\end{table}
\vspace{-10pt}

% ================= EBHI =================
\begin{table}[htbp]
\centering
\caption{Segmentation performance on \textbf{EBHI} under DoRA fine-tuning. Mean with 95\% confidence interval.}
\label{tab:ebhi}
\resizebox{\textwidth}{!}{
\scriptsize
%
}
\vspace{-8pt}
\end{table}
\vspace{-10pt}

% ================= GlaS =================
\begin{table}[htbp]
\centering
\caption{Segmentation performance on \textbf{GlaS} under DoRA fine-tuning. Mean with 95\% confidence interval.}
\label{tab:glas}
\resizebox{\textwidth}{!}{
\scriptsize
%
}
\vspace{-8pt}
\end{table}
\vspace{-10pt}

% ================= Janowczyk =================
\begin{table}[htbp]
\centering
\caption{Segmentation performance on \textbf{Janowczyk} under DoRA fine-tuning. Mean with 95\% confidence interval.}
\label{tab:janowczyk}
\resizebox{\textwidth}{!}{
\scriptsize
%
}
\vspace{-8pt}
\end{table}
\vspace{-10pt}

% ================= Lizard =================
\begin{table}[htbp]
\centering
\caption{Segmentation performance on \textbf{Lizard} under DoRA fine-tuning. Mean with 95\% confidence interval.}
\label{tab:lizard}
\resizebox{\textwidth}{!}{
\scriptsize
%
}
\vspace{-8pt}
\end{table}
\vspace{-10pt}

% ================= NuCLS =================
\begin{table}[htbp]
\centering
\caption{Segmentation performance on \textbf{NuCLS} under DoRA fine-tuning. Mean with 95\% confidence interval.}
\label{tab:nucls}
\resizebox{\textwidth}{!}{
\scriptsize
%
}
\vspace{-8pt}
\end{table}
\vspace{-10pt}

% ================= PanNuke =================
\begin{table}[htbp]
\centering
\caption{Segmentation performance on \textbf{PanNuke} under DoRA fine-tuning. Mean with 95\% confidence interval.}
\label{tab:pannuke}
\resizebox{\textwidth}{!}{
\scriptsize
%
}
\vspace{-8pt}
\end{table}
\vspace{-10pt}

% ================= RINGS =================
\begin{table}[htbp]
\centering
\caption{Segmentation performance on \textbf{RINGS} under DoRA fine-tuning. Mean with 95\% confidence interval.}
\label{tab:rings}
\resizebox{\textwidth}{!}{
\scriptsize
%
}
\vspace{-8pt}
\end{table}
\vspace{-10pt}

% ================= TNBC =================
\begin{table}[htbp]
\centering
\caption{Segmentation performance on \textbf{TNBC} under DoRA fine-tuning. Mean with 95\% confidence interval.}
\label{tab:tnbc}
\resizebox{\textwidth}{!}{
\scriptsize
%
}
\vspace{-8pt}
\end{table}
\vspace{-10pt}

% ================= WSSS4LUAD =================
\begin{table}[htbp]
\centering
\caption{Segmentation performance on \textbf{WSSS4LUAD} under DoRA fine-tuning. Mean with 95\% confidence interval.}
\label{tab:wsss4luad}
\resizebox{\textwidth}{!}{
\scriptsize
%
}
\vspace{-8pt}
\end{table}
\vspace{-10pt}

% ================= cpm15 =================
\begin{table}[htbp]
\centering
\caption{Segmentation performance on \textbf{cpm15} under DoRA fine-tuning. Mean with 95\% confidence interval.}
\label{tab:cpm15}
\resizebox{\textwidth}{!}{
\scriptsize
%
}
\vspace{-8pt}
\end{table}
\vspace{-10pt}

% ================= cpm17 =================
\begin{table}[htbp]
\centering
\caption{Segmentation performance on \textbf{cpm17} under DoRA fine-tuning. Mean with 95\% confidence interval.}
\label{tab:cpm17}
\resizebox{\textwidth}{!}{
\scriptsize
%
}
\vspace{-8pt}
\end{table}
\vspace{-10pt}

% ================= kumar =================
\begin{table}[htbp]
\centering
\caption{Segmentation performance on \textbf{kumar} under DoRA fine-tuning. Mean with 95\% confidence interval.}
\label{tab:kumar}
\resizebox{\textwidth}{!}{
\scriptsize
%
}
\vspace{-8pt}
\end{table}
\vspace{-10pt}

\clearpage
\section{CNN Adapter Performance }
\renewcommand{\thefigure}{F\arabic{figure}}
\renewcommand{\thetable}{F\arabic{table}}
\setcounter{figure}{0}
\setcounter{table}{0}
% ================= BCSS =================
\begin{table}[htbp]
\centering
\caption{Segmentation performance on \textbf{BCSS} under CNN adapter. Mean with 95\% confidence interval.}
\label{tab:bcss}
\resizebox{\textwidth}{!}{
\scriptsize
%
}
\vspace{-8pt}
\end{table}
\vspace{-10pt}

% ================= COSAS24 =================
\begin{table}[htbp]
\centering
\caption{Segmentation performance on \textbf{COSAS24} under CNN adapter. Mean with 95\% confidence interval.}
\label{tab:cosas24}
\resizebox{\textwidth}{!}{
\scriptsize
%
}
\vspace{-8pt}
\end{table}
\vspace{-10pt}

% ================= CRAG =================
\begin{table}[htbp]
\centering
\caption{Segmentation performance on \textbf{CRAG} under CNN adapter. Mean with 95\% confidence interval.}
\label{tab:crag}
\resizebox{\textwidth}{!}{
\scriptsize
%
}
\vspace{-8pt}
\end{table}
\vspace{-10pt}

% ================= CoCaHis =================
\begin{table}[htbp]
\centering
\caption{Segmentation performance on \textbf{CoCaHis} under CNN adapter. Mean with 95\% confidence interval.}
\label{tab:cocahis}
\resizebox{\textwidth}{!}{
\scriptsize
%
}
\vspace{-8pt}
\end{table}
\vspace{-10pt}

% ================= CoNIC2022 =================
\begin{table}[htbp]
\centering
\caption{Segmentation performance on \textbf{CoNIC2022} under CNN adapter. Mean with 95\% confidence interval.}
\label{tab:conic2022}
\resizebox{\textwidth}{!}{
\scriptsize
%
}
\vspace{-8pt}
\end{table}
\vspace{-10pt}

% ================= CoNSeP =================
\begin{table}[htbp]
\centering
\caption{Segmentation performance on \textbf{CoNSeP} under CNN adapter. Mean with 95\% confidence interval.}
\label{tab:consep}
\resizebox{\textwidth}{!}{
\scriptsize
%
}
\vspace{-8pt}
\end{table}
\vspace{-10pt}

% ================= EBHI =================
\begin{table}[htbp]
\centering
\caption{Segmentation performance on \textbf{EBHI} under CNN adapter. Mean with 95\% confidence interval.}
\label{tab:ebhi}
\resizebox{\textwidth}{!}{
\scriptsize
%
}
\vspace{-8pt}
\end{table}
\vspace{-10pt}

% ================= GlaS =================
\begin{table}[htbp]
\centering
\caption{Segmentation performance on \textbf{GlaS} under CNN adapter. Mean with 95\% confidence interval.}
\label{tab:glas}
\resizebox{\textwidth}{!}{
\scriptsize
%
}
\vspace{-8pt}
\end{table}
\vspace{-10pt}

% ================= Janowczyk =================
\begin{table}[htbp]
\centering
\caption{Segmentation performance on \textbf{Janowczyk} under CNN adapter. Mean with 95\% confidence interval.}
\label{tab:janowczyk}
\resizebox{\textwidth}{!}{
\scriptsize
%
}
\vspace{-8pt}
\end{table}
\vspace{-10pt}

% ================= Lizard =================
\begin{table}[htbp]
\centering
\caption{Segmentation performance on \textbf{Lizard} under CNN adapter. Mean with 95\% confidence interval.}
\label{tab:lizard}
\resizebox{\textwidth}{!}{
\scriptsize
%
}
\vspace{-8pt}
\end{table}
\vspace{-10pt}

% ================= NuCLS =================
\begin{table}[htbp]
\centering
\caption{Segmentation performance on \textbf{NuCLS} under CNN adapter. Mean with 95\% confidence interval.}
\label{tab:nucls}
\resizebox{\textwidth}{!}{
\scriptsize
%
}
\vspace{-8pt}
\end{table}
\vspace{-10pt}

% ================= PanNuke =================
\begin{table}[htbp]
\centering
\caption{Segmentation performance on \textbf{PanNuke} under CNN adapter. Mean with 95\% confidence interval.}
\label{tab:pannuke}
\resizebox{\textwidth}{!}{
\scriptsize
%
}
\vspace{-8pt}
\end{table}
\vspace{-10pt}

% ================= RINGS =================
\begin{table}[htbp]
\centering
\caption{Segmentation performance on \textbf{RINGS} under CNN adapter. Mean with 95\% confidence interval.}
\label{tab:rings}
\resizebox{\textwidth}{!}{
\scriptsize
%
}
\vspace{-8pt}
\end{table}
\vspace{-10pt}

% ================= TNBC =================
\begin{table}[htbp]
\centering
\caption{Segmentation performance on \textbf{TNBC} under CNN adapter. Mean with 95\% confidence interval.}
\label{tab:tnbc}
\resizebox{\textwidth}{!}{
\scriptsize
%
}
\vspace{-8pt}
\end{table}
\vspace{-10pt}

% ================= WSSS4LUAD =================
\begin{table}[htbp]
\centering
\caption{Segmentation performance on \textbf{WSSS4LUAD} under CNN adapter. Mean with 95\% confidence interval.}
\label{tab:wsss4luad}
\resizebox{\textwidth}{!}{
\scriptsize
%
}
\vspace{-8pt}
\end{table}
\vspace{-10pt}

% ================= cpm15 =================
\begin{table}[htbp]
\centering
\caption{Segmentation performance on \textbf{cpm15} under CNN adapter. Mean with 95\% confidence interval.}
\label{tab:cpm15}
\resizebox{\textwidth}{!}{
\scriptsize
%
}
\vspace{-8pt}
\end{table}
\vspace{-10pt}

% ================= cpm17 =================
\begin{table}[htbp]
\centering
\caption{Segmentation performance on \textbf{cpm17} under CNN adapter. Mean with 95\% confidence interval.}
\label{tab:cpm17}
\resizebox{\textwidth}{!}{
\scriptsize
%
}
\vspace{-8pt}
\end{table}
\vspace{-10pt}

% ================= kumar =================
\begin{table}[htbp]
\centering
\caption{Segmentation performance on \textbf{kumar} under CNN adapter. Mean with 95\% confidence interval.}
\label{tab:kumar}
\resizebox{\textwidth}{!}{
\scriptsize
%
}
\vspace{-8pt}
\end{table}
\vspace{-10pt}

\clearpage 
\section{Transformer Adapter Performance }
\renewcommand{\thefigure}{G\arabic{figure}}
\renewcommand{\thetable}{G\arabic{table}}
\setcounter{figure}{0}
\setcounter{table}{0}
% ================= BCSS =================
\begin{table}[htbp]
\centering
\caption{Segmentation performance on \textbf{BCSS} under Transformer adapter. Mean with 95\% confidence interval.}
\label{tab:bcss}
\resizebox{\textwidth}{!}{
\scriptsize
%
}
\vspace{-8pt}
\end{table}
\vspace{-10pt}

\end{document}